\documentclass[manuscript]{aastex63}
\usepackage{natbib}
\usepackage{graphicx}
\usepackage{bmpsize}
\usepackage{enumerate}
\submitjournal{ApJ}

\shorttitle{twin outflow from vla 1623}
\shortauthors{Hara et al.}

\begin{document}

\title{Misaligned Twin Molecular Outflows From Class-0 Proto-stellar Binary System VLA 1623A Unveiled by ALMA}

\correspondingauthor{Chihomi Hara}
\email{c.hara}

\author{Chihomi Hara}
\affiliation{NEC Coorporation\\
 Fuchu, Tokyo 183-8501, Japan}
\affiliation{ Department of Astronomy, The University of Tokyo, \\
7-3-1, Hongo, Bunkyo-ku, Tokyo, 113-0033, Japan}

\author[0000-0002-8049-7525]{Ryohei Kawabe}
\affiliation{ Department of Astronomy, The University of Tokyo, \\
7-3-1, Hongo, Bunkyo-ku, Tokyo, 113-0033, Japan}
\affiliation{National Astronomical Observatory of Japan, \\
Mitaka, Tokyo 181-8588, Japan}
\affiliation{The Graduate University for Advanced Studies (SOKENDAI), \\
2-21-1 Osawa, Mitaka, Tokyo 181-8588, Japan}

\author[0000-0001-5431-2294]{Fumitaka Nakamura}
\affiliation{National Astronomical Observatory of Japan, \\
Mitaka, Tokyo 181-8588, Japan}
\affiliation{The Graduate University for Advanced Studies (SOKENDAI), \\
2-21-1 Osawa, Mitaka, Tokyo 181-8588, Japan}
\affiliation{ Department of Astronomy, The University of Tokyo, \\
7-3-1, Hongo, Bunkyo-ku, Tokyo, 113-0033, Japan}

\author[0000-0001-9304-7884]{Naomi Hirano}
\affiliation{Academia Sinica Institute of Astronomy and Astrophysics, \\ 
11F of Astronomy-Mathematics Building, AS/NTU, No.1, Sec. 4, Roosevelt Rd, Taipei, 10617, Taiwan}

\author[0000-0003-0845-128X]{Shigehisa Takakuwa}
\affiliation{Department of Physics and Astronomy, Graduate School of Science and Engineering, Kagoshima University, \\
1-21-35 Korimoto, Kagoshima, Kagoshima 890-0065, Japan}

\author[0000-0001-9368-3143]{Yoshito Shimajiri}
\affiliation{National Astronomical Observatory of Japan, \\
Mitaka, Tokyo 181-8588, Japan}

\author[0000-0002-2067-629X]{Takeshi Kamazaki}
\affiliation{National Astronomical Observatory of Japan, \\
Mitaka, Tokyo 181-8588, Japan}

\author[0000-0002-9289-2450]{James Di Francesco}
\affiliation{NRC Herzberg Inst of Astrophysics, \\
5071 W Saanich Rd, Victoria BC V9E 2E7, British Columbia, Canada}

\author[0000-0002-0963-0872]{Masahiro N. Machida}
\affiliation{Department of Earth and Planetary Science, Faculty of Science, Kyushu
University, \\
Motooka 744 , Nishi-ku, Fukuoka 819-0395, Japan}

\author[0000-0002-6510-0681]{Motohide Tamura}
\affiliation{National Institutes of Natural Sciences, Astrobiology Center, \\
2-21-1, Osawa, Mitaka, Tokyo 181-8588, Japan}
\affiliation{ Department of Astronomy, The University of Tokyo, \\
7-3-1, Hongo, Bunkyo-ku, Tokyo, 113-0033, Japan}

\author[0000-0003-1549-6435]{Kazuya Saigo}
\affiliation{National Astronomical Observatory of Japan, \\
Mitaka, Tokyo 181-8588, Japan}

\author[0000-0002-8125-4509]{Tomoaki Matsumoto}
\affiliation{Faculty of Sustainability Studies, Hosei University, \\
Fujimi, Chiyoda-ku, Tokyo 102-8160, Japan}

\author[0000-0001-8105-8113]{Kengo Tomida}
\affiliation{Department of Earth \& Space Science, Osaka University, \\
Osaka 560-0043, Japan}
\affiliation{Astronomical Institute, Tohoku University, Sendai, Miyagi 980-8578, Japan}



\begin{abstract}
We present the results of ALMA observations toward the low-mass Class-0 binary system, VLA 1623Aab in the Ophiuchus molecular cloud in $^{12}$CO, $^{13}$CO, and C$^{18}$O(2--1) lines.
Our $^{12}$CO ($J$=2--1) data reveal that the VLA 1623 outflow  consists of twin spatially overlapped outflows/jets.
The redshifted northwestern jet exhibits the three cycles of wiggle with a spatial period of 1360$\pm$10 au, corresponding to a time period of 180 yr. The wiggle-like structure is also found in the position-velocity (PV) diagram, showing an amplitude in velocity of about 0.9 km s$^{-1}$. Both the period and the velocity amplitude of the wiggle are roughly consistent with those expected from the binary parameters, i.e., the orbital period (460$\pm$20 yr) and the Keplerian velocity (2.2 km s$^{-1}$). Our $^{13}$CO and C$^{18}$O images reveal the nature of the dense gas in the two cm/mm sources, VLA 1623-B and -W, and its relation to the outflows, and strongly support the previous interpretation that both are shocked cloudlets. The driving sources of the twin molecular outflows are, therefore, likely to be within the VLA 1623Aab binary. The axes of the two molecular outflows are estimated to be inclined by 70$\arcdeg$ from each other across the plane of sky, implying that the associated protostellar disks are also misaligned by $70\arcdeg$. Such a misalignment, together with a small binary separation of 34 au in the one of the youngest protobinary systems known, is difficult to explain by models of disk fragmentation in quiescent environments. Instead, other effects such as turbulence probably play roles in misaligning the disks.
\end{abstract}



\section{Introduction} \label{sec:intro}
\subsection{binary formation}
A large fraction of low-mass stars are born in binary systems \citep {leinert93,raghavan10}. Therefore, investigating how binary systems form is crucial to understanding low-mass star formation. Several theoretical scenarios of binary formation have been suggested including the fragmentation of a gravitationally unstable disk (disk fragmentation model; \citealt{adams89,bonnell94}) and the fragmentation of a core (the model without turbulence; \citealt{matsumoto03}, turbulent fragmentation model; \citealt{offner10}, and formation of binaries in a cluster ; \citealt{bate13}). Although several observational studies have investigated which mechanism is dominant \citep[e.g.,][]{tobin18}, the debate is not yet settled.

Previous theoretical studies have indicated that the rotational axes of protostellar disks in a binary system can provide a key to understanding the formation process of binaries. For example, the classical numerical simulations of disk fragmentation predict that the rotational axes of the protostellar disks should be parallel to that of the circum-binary disk \citep{adams89,bonnell94}.
In contrast, other models such as turbulent fragmentation and capture suggest that the rotational axes can be misaligned \citep{offner10,bate10,bate13}. \cite{bate00} also posited that misaligned disks should be brought into rough alignment by tidal torques in 20 binary orbital periods. Hence, we need to study young proto-binary systems.

Several observational studies have attempted to investigate the alignments of disks from jet/outflow orientations since jets are considered to be launched perpendicular to disks. These studies, however, have been limited to relatively wide binary systems whose separations involve several thousand au \citep{lee16}. Observations of  molecular outflows from binary systems can therefore help us to reveal the alignment of the disks in binaries and probe the mechanism of their formation.

In recent studies, an aligned binary was discovered in the L1448 IRS 3 system making it consistent with the classical disk fragmentation scenario \citep{tobin16}, or the alignment by tidal torques.
In contrast, the IRS 43 and L1551 NE systems observed by \cite{brinch16} and \cite{takakuwa17}, respectively, are found to be misaligned. Thus, the dominant binary formation mechanism continues to be a matter of debate. In thie paper, we examine at high resolution, another binary, VLA 1623A.   


\subsection{VLA 1623}
 VLA 16234-2417 (hereafter VLA 1623) is the prototypical Class 0 object \citep{Andre90}. It produces 
the most energetic outflow in the Oph A core \citep{Andre90,dent95,nakamura11,white14}, which itself is located in the Ophiuchus star-forming region at a distance of 120--140 pc \citep{knude98,loinard08,lombardi08,mamajek08}.
 In this study, we adopt a distance to VLA 1623 of 137.3$\pm$1.2 pc \citep{ortiz17}, since it is one of the most recent values obtained with Very Long Baseline Array (VLBA) observations and it falls into the range of the former measurements, 120--140 pc. 

\begin{figure}
\centering
\includegraphics[scale=0.33]{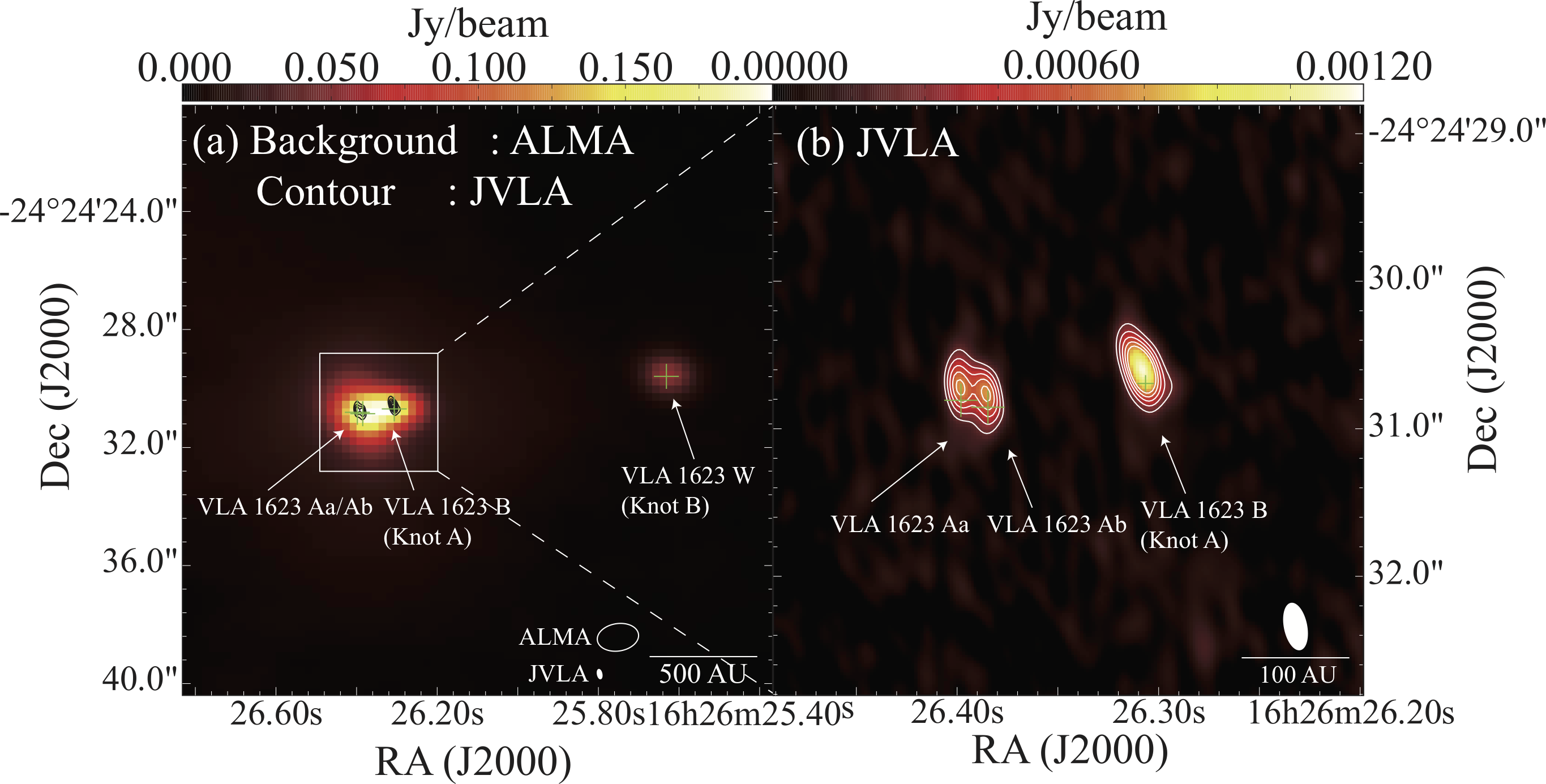}
\caption{The continuum maps of VLA 1623. The contours in (a) and (b) show the 7.3 mm continuum emission obtained with The Karl G. Jansky Very Large Array (JVLA) observations in \citep{kawabe18}. The contour levels start from 6 $\sigma$ with the interval of 3 $\sigma$ until 21 $\sigma$ (1 $\sigma$=0.038 mJy beam$^{-1}$). The background colors in (a) and (b) show the ALMA 1.3 mm continuum image and JVLA 7.3 mm continuum image, respectively. Crosses show the positions of VLA 1623Aab/B/W obtained by \cite{harris18}.}
\label{jvla}
\end{figure}

Previous observations of VLA 1623 have revealed 
a highly complex structure in the protostellar system, as shown in Figure \ref{jvla}, and its physical characteristics are still controversial.
From VLA observations, \citet{bontemps97} identified four compact cm sources : VLA 1623A, VLA 1623B (Knot A), VLA 1623W (Knot B), and Knot C.  
Later, \citet{looney00}, based on the Berkeley Illinois Maryland Association (BIMA) observations, proposed that the VLA 1623A/B is a binary system. 
More recently, \cite{maury12} suggested that VLA 1623B/W are cloudlets created by outflow shocks.
\cite{murilo13}, however, supported \cite{looney00}'s interpretation and claimed that VLA 1623B/W are Class-0 and Class-I protostars based on their spectral energy distributions (SEDs).
Interestingly, the other source, Knot C, was detected on the northwest of VLA 1623W in deep VLA observations \citep{bontemps97}, but was not detected in the Gould's Belt Very Large Array Survey \citep{Dzib13}.

In more recent observations, VLA 1623A itself was revealed to be consisting of VLA 1623Aa and VLA 1623Ab \citep{harris18,kawabe18}. VLA 1623Aab are surrounded by the Keplerian circum-binary disk (hereafter CB disk) with a radius of 150--180 au\citep{murilo13,heish19}.
Thus, VLA 1623 is an excellent target to study the initial state of the binary system, e.g., by measuring the relative angle of the outflow axes.

The molecular outflow from VLA 1623A has blueshifted and redshifted lobes on both the northwest and southeast of VLA 1623 \citep{Andre90,dent95}. \cite{Andre90} concluded that a single biconical flow is being ejected from VLA 1623A/B. \cite{dent95}, however, suggested that two distinct molecular outflows are being ejected from VLA 1623A/B since two shocked H$_2$ spots are located in parallel on its northwest. 

In this study, we present our 1$\arcsec$ resolution (corresponding to 137 au) ALMA observations in $^{12}$CO, $^{13}$CO, and C$^{18}$O(2--1) lines toward VLA 1623. First, we describe the details of our ALMA observations in Section \ref{sec:obs}. In Section \ref{sec:res}, we show the results of ALMA observations with which we successfully resolved the two outflows associated with VLA 1623A, and we derive their physical properties. In Section \ref{sec:dis}, we discuss the origin of the two outflows and attempt to understand the formation of the binary system. Finally, we summarize our main conclusions in Section \ref{sec:con}.

\section{Observations} \label{sec:obs}
\label{sec:obs:alma}
We observed 1.3-mm continuum emission and $^{12}$CO($J=$2--1; 230.538 GHz), $^{13}$CO($J=$2--1; 220.399 GHz), and C$^{18}$O($J=$2--1; 219.560 GHz) emission toward the Oph A core using ALMA's 12-m Array and 7-m Array (PI: F. Nakamura; Project code: 2013.1.00839.S). In this study, we focus on the results of the molecular line observations in the vicinity of VLA 1623. The images of the continuum and molecular lines for the whole area observed will be presented in the forthcoming paper (Hara et al., in prep), where we will also discuss the molecular outflows and dense gas structures associated with other protostellar sources in the area, such as GSS30-IRS1 and -IRS3. We note that a part of the 1.3-mm dust continuum map was already presented in \cite{kawabe18}.  

 The observations span an area of 150$\arcsec$ (in RA) $\times$180$\arcsec$ (in decl.)  with 150 12-m Array pointings and 50 7-m Array pointings, with spacings of 13\farcs6 and 23\farcs2, respectively. In this study, we focus on the part of our ALMA Oph A 150$\arcsec\times120\arcsec$ that contains the  molecular outflow from VLA 1623A/B. In our observations, the phase reference center of both dataset was set to be ($\alpha_{\rm J2000.0}$,$\delta_{\rm J2000.0}$)=(16$^{\rm h}$27$^{\rm m}$26\fs507, -24\arcdeg31\arcmin28\farcs63), which is the same as that of Oph-B2 observations \citep{kamazaki19}.

The observations were performed on May 1st, 2015, with the 12-m Array and in the period May 14th to 17th, 2015 with the 7-m Array. The $uv$ ranges sampled in these datasets are 12.5 k$\lambda$--348 k$\lambda$ and 8.1 k$\lambda$--48 k$\lambda$, respectively. The minimum $uv$ distance of the combined data corresponds to 25\arcsec. Observations of the calibrator J1625-2527 were interleaved with the target for complex gain calibration in both datasets. The passband response of the 12-m Array and 7-m Array data was calibrated using a strong source, J1516-2422 and J1733-1304, respectively. Absolute flux scales of 12-m Array and 7-m Array data were determined with Titan and Mars, respectively. The data were calibrated and combined using the Common Astronomy Software Applications package, CASA pipeline version 4.2.2 and version 4.5.3, respectively. Images were made and deconvolved using CLEAN method with the natural $uv$ weighting.

For the $^{12}$CO data, the resulting synthesized beam size was 1\farcs39$\times$0\farcs91 (corresponding to 190 au$\times$125 au) with a position angle of 89\arcdeg. The achieved rms noise level is measured to be 15 mJy beam$^{-1}$ for a map with a velocity resolution of 1.0 km s$^{-1}$. The 0.2 km s$^{-1}$ resolution maps archive noise levels of 50 mJy beam$^{-1}$, 34 mJy beam$^{-1}$, and 28 mJy beam$^{-1}$ in the $^{12}$CO, $^{13}$CO and C$^{18}$O(2--1) lines, respectively. We mainly used the $^{12}$CO map with a velocity grid of 1.0 km s$^{-1}$ to investigate the overall velocity structure of the outflows. We also used the $^{12}$CO map with a velocity grid of 0.2 km s$^{-1}$ to investigate the velocity structure in detail. The observations are summarized in Table. \ref{tbl:obs}.


\section{Results} \label{sec:res}
\subsection{Two outflows from VLA 1623} \label{sec:res:co1}

\begin{figure}
\centering
\includegraphics[scale=0.6]{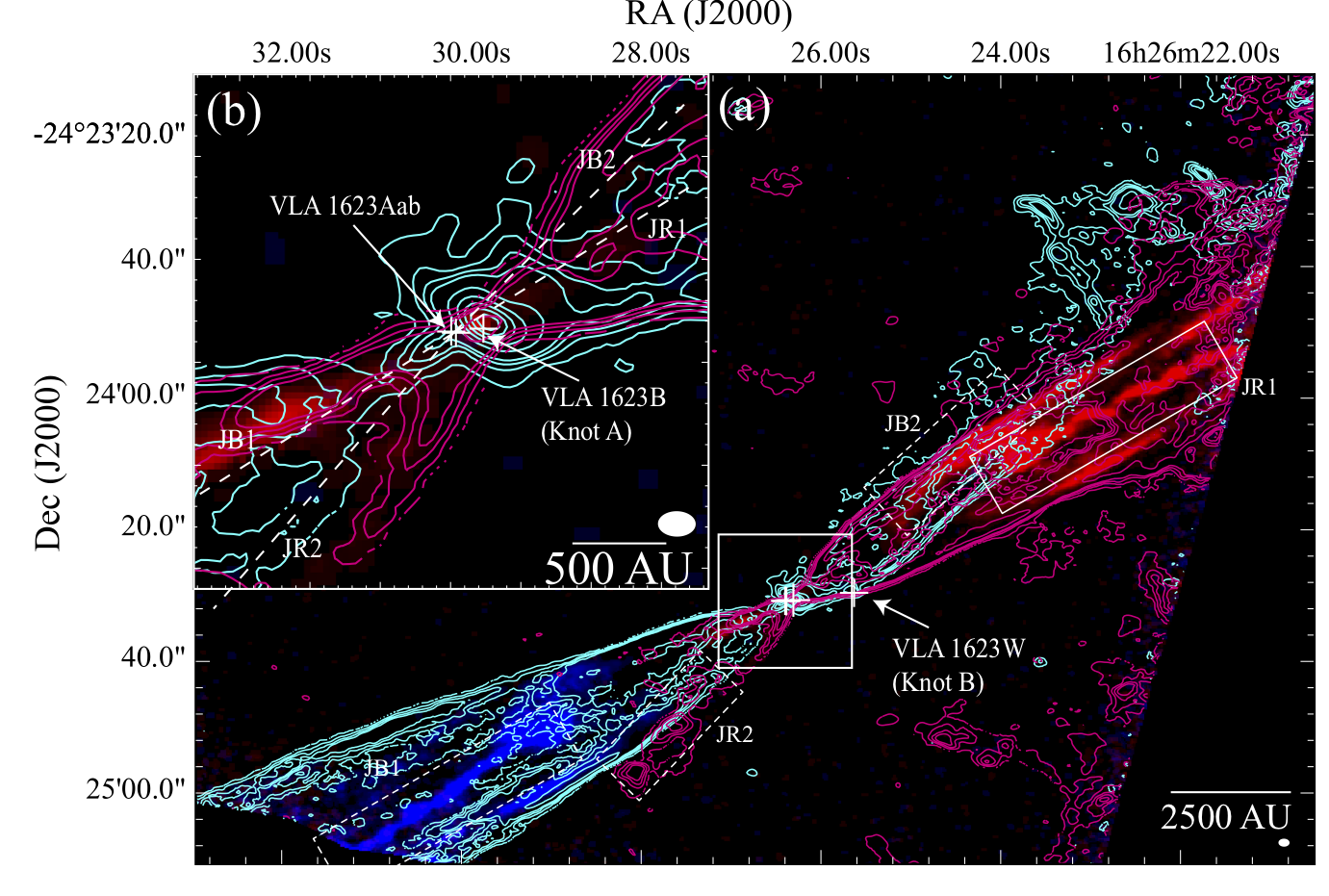}
\caption{Overlay of the low-velocity $^{12}$CO(2--1) on the high-velocity $^{12}$CO maps. The blue and red colors in (a) and (b) represent blueshifted and redshifted intensities integrated between $V_{\rm LSR}$=-34.5 to -10.5 km s$^{-1}$ and 14.5 to 31.5 km s$^{-1}$. The cyan and magenta contours superposed on 2-color image in (a) and (b) indicate the emission integrated over intervals from $V_{\rm LSR}$=-10.5 to -0.5 km s$^{-1}$, and from 6.5 to 14.5 km s$^{-1}$. Contour levels start at 5$\sigma$ noise level and increase with an interval of 10$\sigma$ (1$\sigma$=0.05 Jy beam$^{-1}$ km s$^{-1}$ for the blueshifted emission and 0.04 Jy beam$^{-1}\cdot$km s$^{-1}$ for the redshifted emission, respectively) until 35 $\sigma$. After 35 $\sigma$, the contour intervals are 40 $\sigma$. The crosses show the positions of VLA 1623A/B/W. The positions of VLA 1623A/B are obtained from \cite{harris18}, and positions of the other Class-0/I and Class-II sources are taken from \cite{evans09}, \cite{gutermuth09}, \cite{friesen14}, and \cite{kawabe18}. The solid white box on (a) shows the area of (b) and the area of Figure \ref{jr1} (a). Dashed boxes in (a) show the jets (JB1, JR1, JB2, and JR2) identified in Section \ref{sec:res:co1}, and dashed lines in (b) show the directions of each jet identified in Section \ref{sec:res:coch}.}
\label{comom}
\end{figure}

\begin{figure}
\centering
\includegraphics[scale=0.6]{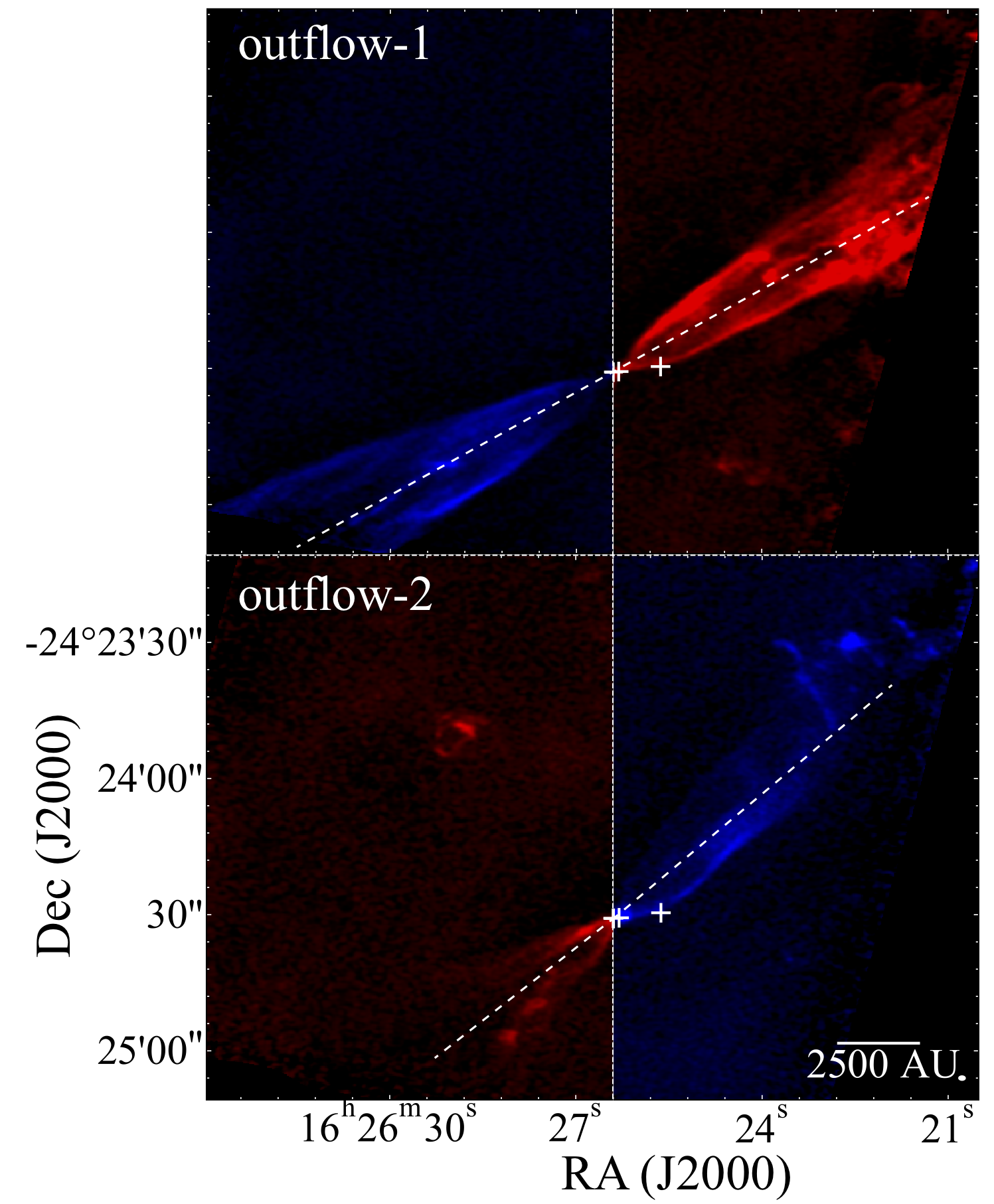}
\caption{$^{12}$CO maps of outflow-1 and outflow-2. The top-left and bottom-right panels (blue colors) represent blueshifted intensities integrated between $V_{\rm LSR}$=-34.5 and -0.5 km s$^{-1}$. The top-right and bottom-left panels (red colors) represent redshifted intensities integrated between $V_{\rm LSR}$=6.5 and 31.5 km s$^{-1}$. The left and right panels show the east and west of VLA 1623A, respectively. The symbols are the same as those in Figure \ref{comom}. The white dashed line depicts the PAs of each shell summarized in Table \ref{tbl:co}.}
\label{comom_sep}
\end{figure}

The obtained $^{12}$CO(2--1) results are presented in Figures \ref{comom} and \ref{comom_sep}. As shown in Figure \ref{comom}, blueshifted and redshifted components are seen in both the northwest and southeast of VLA 1623A/B. Although the overall structure is consistent with previous studies \citep{Andre90,dent95,yu97}, the outflow has been resolved into twin molecular outflows from the vicinity of VLA 1623A/B as shown in Figure \ref{comom_sep}. 
 
The top panels of Figure \ref{comom_sep} show the southeastern blue lobe and the northwestern red lobe pair, while the bottom panels show the southeastern red lobe and the northeastern blue lobe pair. We assume the velocity range of the molecular cloud is from $V_{\rm LSR}=$-0.5 to $+$6.5 km s$^{-1}$ since the systemic velocity of VLA 1623A is $V_{\rm LSR}=$4 km s$^{-1}$ from the fitting of the CB disk \citep{murilo13}. The axes of the outflow pairs presented in the top and bottom panels are different by roughly 10$\arcdeg$, which are measured by eye. Therefore, we interpret that there are two pairs of outflows being ejected from the vicinity of VLA 1623A/B, as suggested previously by \cite{dent95}, rather than a single outflow in an edge-on configuration \citep{Andre90}. As described later in Section \ref{sec:res:coch}, each of the four lobes is associated with its own jet-like structure, supporting the existence of two molecular outflows. Indeed, the measured inclination angle of the CB disk and circumstellar disks (hereafter, CS disks), 45\arcdeg--55$\arcdeg$ \citep{murilo13,harris18}, is incompatible with an edge-on outflow, where the CB disk or CS disks needs to be $>$80$\arcdeg$ \citep{Andre90}. 

Hereafter, we refer to the outflow shown in the top panel of Figure \ref{comom_sep}  ``outflow-1'', and that in the bottom panel ``outflow-2''. We also refer to the blueshifted and redshifted outflow shells (lobes) in outflow-1 as SB1 and SR1, and those in outflow-2 as SB2 and SR2, respectively. The position angles of the four outflow shell axes, which are roughly measured by eye, are listed in Table \ref{tbl:co}.  

\subsection{$^{12}$CO velocity channel maps} \label{sec:res:coch}

In this subsection, we analyze the details of outflow-1 and outflow-2 in Figures \ref{coch} (channel maps) and \ref{coch_center} (close-up of channel maps). We find collimated jet-like structures associated with all four outflow whells (SB1, SR1, SB2, and SR2), as shown in Figure \ref{jets}. Hereafter, we refer to the collimated jet-like structures inside the outflow lobes as jets. 

The main features in Figures \ref{coch} and \ref{coch_center} are summarized below;
\begin{enumerate}[a{)}]
\item (SB1 and SR1) As already shown in Section \ref{sec:res:co1}, the blueshifted outflow shell, which corresponds to SB1, is observed in the channels corresponding to $V_{\rm LSR}=$-18.0 to -3.0 km s$^{-1}$. In contrast, SR1 is detected in the channels corresponding to $V_{\rm LSR}=+$12.0 to $+$27.0 km s$^{-1}$. 

\item (JB1) There is a blueshifted jet (hereafter JB1) in the velocity channels corresponding to $V_{\rm LSR}=$-33.0 to -13.0 km s$^{-1}$. JB1 consists of two features in the $V_{\rm LSR}=$-18.0 km s$^{-1}$ channel. An integrated intensity map of JB1 is shown in Figure \ref{jets} (a). The position angle (PA) of JB1 is approximately 120\arcdeg. The axis of JB1 is almost the same as that of SB1 (outflow-1). Therefore, JB1 is considered to be associated with outflow-1.

\item (JR1) In the opposite direction of JB1, a redshifted jet-like structure, JR1, is observed in the velocity channels between $V_{\rm LSR}=+$17.0 km s$^{-1}$ and $V_{\rm LSR}=+$32.0 km s$^{-1}$. JR1 is considered to be associated with SR1, which is observed in the velocity channel range of $V_{\rm LSR}=+$7.0 to $+$27.0 km s$^{-1}$, since the directions are the same. The integrated intensity map of JR1 is shown in Figure \ref{jets} (c). 

\item (SB2 and SR2) The blueshifted outflow shell associated with outflow-2 (SB2) is observed in the velocity channel range of $V_{\rm LSR}=$-8.0 to -3.0 km s$^{-1}$. In the opposite direction of SB2, SR2 is in the velocity channel range of $V_{\rm LSR}=+$12.0 to $+$22.0 km s$^{-1}$. In the magnified view (Figure \ref{coch_center}), the redshifted outflow shell on the southeastern side of VLA 1623  in the velocity channel range of $V_{\rm LSR}=+$12.0 to $+$22.0 km s$^{-1}$, which is considered to be SR2, is elongated in a direction different from that of JR1.

\item (JB2) There is a jet-like structure in the SB2 outflow shell in the velocity channel range of $V_{\rm LSR}=$-3.0 to -8.0 km s$^{-1}$. We call this jet-like structure JB2. JB2 is weakly detected in the channel maps but is clearer to see in the integrated intensity map, as shown in Figure \ref{jets} (d). 

\item (JR2) A compact condensation inside SR2 is found in the velocity channel of $V_{\rm LSR}=+$12.0 km s$^{-1}$. This feature seems to be a part of jet-like structure in the $V_{\rm LSR}=+7.0$ km s$^{-1}$ channel and the integrated intensity map of Figure \ref{jets} (b). Hereafter, we refer to this jet-like structure as JR2. JR2 is found to be composed of four condensations, as shown in Figure \ref{jets} (b).

\item (JB1-2 and JR1-2) Taking a closer look at the center of VLA 1623A/B (Figure \ref{coch_center}), the blueshifted structure in the $V_{\rm LSR}=$-8.0 km s$^{-1}$ channel is elongated from VLA 1623A in the same direction as JB1. This structure is denoted as JB1-2 in Figure \ref{coch_center}. In contrast, a redshifted structure in the channels corresponding to $V_{\rm LSR}=+$17.0 to $+$22.0 km s$^{-1}$ in Figure \ref{coch_center} is elongated from the VLA 1623A/B region in the same direction as JR1. We label this structure as JR1-2 in Figure \ref{coch_center}. JB1-2 and JR1-2 are considered to be also associated with outflow-1 since their axes are similar to JB1/SB1 and JR1/SR1, respectively. 

\item (compact feature on VLA 1623B) A compact blueshifted component is associated with the northwestern side of  VLA 1623B in the velocity channel range of $V_{\rm LSR}=$-3.0 to $+$2.0 km s$^{-1}$ (see Figure \ref{coch_center}). 

\end{enumerate}

The features are also described in Appendix \ref{sec:app:chan}. JB1 and JR1 are considered to be associated with outflow-1, while JB2 and JR2 are considered to be associated with outflow-2. The velocities and position angles of the jets and shells are summarized in Table \ref{tbl:co}. Table \ref{tbl:co} also lists the opening angles of each outflow lobe, as measured by eye. We cannot calculate the dynamical time of the outflow shells, since the map does not span their entire extents. We do, however, calculate the dynamical time of jets. Since the jets are possibly elongated beyond the map, these dynamical times of jets marks the lower limits. Moreover, we note that the axes of JR1 and JB1, which are associated with outflow-1, are perpendicular to the CB disk PA of 30\arcdeg--35$\arcdeg$\citep{murilo13,heish19}. Therefore, outflow-1 is considered to be driven by VLA 1623A. 

\subsection{Outflow Velocity Structure}
\label{sec:res:copv}
In this subsection, we investigate the internal structures of the jets and outflow shells using the position-velocity (PV) diagrams in Figures \ref{copv} and \ref{copv2}, and the integrated intensity maps in Figure \ref{subshells}. Figure \ref{copv} (a) and Figure \ref{copv2} (a) show PV diagrams cut along the jets associated with outflow-1 and outflow-2, respectively. In Figure \ref{copv} (a), several velocity components exhibit mostly linear increases of velocity along the spatial axis, indicating that the corresponding jets and shells are accelerated. 

In the redshifted side, the highest velocity component is JR1. It is wiggling along the velocity axis, as shown in the magnified view of JR1 in Figure \ref{jr1} (b). Also in the redshifted side of Figure \ref{copv} (a), two further components are accelerated from VLA 1623A/B and a position 56$\arcsec$ away from VLA 1623A/B, i.e., SR1 and SR1-2. For the latter, we create the integrated intensity map shown in Figure \ref{subshells} (d). A shell larger than SR1 exists across SR1 on its northwestern side. In the blueshifted side, JB1 is accelerated from a position 45$\arcsec$ away from VLA 1623A/B, and SB1 is accelerated from VLA 1623A/B themselves. At the starting point of JB1, a compact feature is found with a large velocity width. This feature does not appear to be associated with JB1, since that feature is elongated more toward the east compared to JB1 in Figure \ref{jets} (a).

The magnified view of Figure \ref{copv} (a) is shown in Figure \ref{copv} (b). There is both redshifted and blueshifted compact structures at the northwestern side of VLA 1623A. The compact jet-like structure, JR1-2, is accelerated more rapidly compared to SR1 at the northwest of VLA 1623A. At the southeast of VLA 1623A, blueshifted gas is accelerated from VLA 1623A and a position 4$\arcsec$ away from VLA 1623A. The component from VLA 1623A corresponds to JR1-2, as described in Section \ref{sec:res:coch}. 

In Figure \ref{copv2} (a), the low-velocity jet-like structures JR2 and JB2 are marginally detected in the velocity ranges between $V_{\rm LSR}=+$6.5 and $+$13.5 km s$^{-1}$, and $V_{\rm LSR}=$-7.5 and $+$0.5 km s$^{-1}$, respectively.  

Figures \ref{copv2} (b) and (c) illustrate PV diagrams cut along the northern edge of  SB1 and SR1. In Figure \ref{copv2} (b), we see that the blueshifted lobe is accelerated two times. We refer to the component accelerated at a position 43$\arcsec$ away from VLA 1623A as SB1-2. The integrated intensity map of SB1-2 is shown in Figure \ref{subshells} (b). SB1-2 is located across the southeastern side of SB1
. 
In Figure \ref{copv2} (c), the redshifted component is accelerated three times, at VLA 1623A/B and at position 23$\arcsec$ and 64$\arcsec$ away from VLA 1623A/B. From its position, the latter correspond to SR1-2. We refer to the component 23 $\arcsec$ from VLA 1623A/B as SR1-3. It is rapidly accelerated compared to SR1. As shown in Figure \ref{subshells} (e), SR1-3 is marginally detected inside SR1. 

The numerous outflow features described here were likely produced by episodic mass ejections. Such episodic mass ejections have been reported in other Class-0/I protostars (e.g., HH 47 IRS; \citet{arce13}, CARMA-7; \citet{Plunkett15}). In those previous studies, episodic mass ejections were believed to be linked to episodic mass accretions onto the central protostars. The timescale of the mass ejection events in outflow-1 is calculated to be 750-1300 years for the redshifted outflow lobe (SR1), assuming $V_{\rm LSR}$=18 km s$^{-1}$, $V_{\rm sys}=$4 km s$^{-1}$, and an inclination of 55$\arcdeg$. The timescale of episodic mass ejections of the blueshifted outflow (SB1) is calculated to be 1100 years, assuming $V_{\rm LSR}=$-14 km s$^{-1}$ and the same assumptions made for SR1.

\subsection{Wiggle of the jet-like structure}
\label{sec:res:wig}
Figure \ref{jr1} (a) illustrates a closer look at JR1, revealing its wiggling structure in position. A wiggle-like structure in velocity is also observable in Figures \ref{jr1} (c). The images in Figures \ref{jr1} (a) and (c) are created from the maps with a 0.2 km s$^{-1}$ grid to observe the wiggle velocity structures in more detail. In Figure \ref{jr1} (a), the spatial wiggle extends at least three cycles over 29$\arcsec$ (corresponding to 4000 au). The spatial periods of the wiggles in the integrated intensity map (Figure \ref{jr1} (a)) and PV diagram (Figure \ref{jr1} (c)) are roughly the same. 

In Figure \ref{jets} (a), JB1 also seems to be wiggling. This is clearest in the $V_{\rm LSR}=$-18.0 km s$^{-1}$ channel of Figure \ref{coch}. The wiggle in JB1 is, however, not as obvious compared to JR1. Therefore, we mainly discuss the wiggle in JR1 in this study.

\subsection{Outflow Physical Properties} \label{sec:res:phy}
We derived physical properties of each outflow feature assuming an excitation temperature of 30 K for $^{12}$CO(2--1) from the past measurement by \cite{white14} for VLA 1623 outflow. A disk inclination angle of $i_{\rm disk}=$55\arcdeg\hspace{1ex} is used according to \citet{murilo13} as well as Section \ref{sec:res:wig}. We calculate the outflow mass ($M_{\rm out}$), momentum ($P_{\rm out}$), and energy ($E_{\rm out}$) assuming local thermal equilibrium (LTE) and using the following equation from \cite{hara13}; 
 
\begin{eqnarray}
M_{\rm out}&=&\sum_j M_j=\sum_j 8.11\times10^{-7}\left (\frac{X_{^{12}{\rm CO}}}{1.0\times10^{-4}}\right )^{-1}\left (\frac{d}{137{\rm pc}} \right )^2 \nonumber
\\ &\hspace{2ex}& \sum_I\left (\frac{S_{\nu,I,j}}{\rm Jy \cdot km\hspace{1ex}s^{-1}}\right ) \left (\frac{T_{\rm ex}}{30\rm K}\right )\exp{[16.6 {\rm K}/T_{\rm ex}]}\hspace{1ex}M_{\sun} 
\\P_{\rm out}&=&\sum_j M_j (v_j/\sin{i_{\rm disk}})
\\E_{\rm out}&=&\frac{1}{2}\sum_j M_j (v_j/\sin{i_{\rm disk}})^2
\label{eq:co21}
\end{eqnarray}
$M_j$ is the mass included in each channel ($j$), $X_{\rm ^{12}CO}$ is abundance ratio of $^{12}$CO, $d$ is the distance to the star, $S_{\nu,i,j}$ is the flux density at each pixel ($I$) and each channel ($j$), $T_{\rm ex}$ is the excitation temperature, $v_j$ is the line-of-sight velocity difference from the systemic velocity of 4 km s$^{-1}$, and $i_{\rm disk}$ is the disk inclination angle.

 The calculated mass, momentum, and kinetic energy of the outflows are summarized in Table \ref{tbl:co}. Notably, our map does not cover the whole area of these outflows. Hence, all values are the lower limits. Since it is difficult to discriminate jets and outflow lobes, we calculate the physical properties of both. 

As shown in Table \ref{tbl:co}, the momentum and the energy of outflow-1 are about 14 times and 7 times larger than those of outflow-2, respectively. This difference implies that the accretion rate of the outflow-1 driving source is larger compared to that of the outflow-2 driving source.

\subsection{Nature of dense gas around VLA 1623B and VLA 1623W} \label{sec:res:13coc18o}
Figures \ref{13coc18o} (a), (b), (c), and (d) show the total integrated intensity maps of $^{13}$CO and C$^{18}$O. High-velocity outflow components, whose velocity ranges are $V_{\rm LSR}=$-34.5 to -0.5 (blueshifted) and $+$6.5 to $+$31.5 km s$^{-1}$ (redshifted), are overlaid. A compact ($\sim 1\farcs4$ corresponding to 200 au) core in $^{13}$CO and C$^{18}$O emission is seen at the position of VLA 1623W. To the northwest of VLA 1623A, $^{13}$CO emission is strongly detected along the northern edge of blueshifted outflow lobe, SB2. In the C$^{18}$O maps, cavity-like structures are elongated from VLA 1623A/B to the northwest and southeast. These cavity-like structures are extended along the blueshifted outflow lobes, SB1 and SB2. Hence, these structures likely trace the outflow cavity walls created by the blueshifted molecular outflows. The fact that the blueshifted emission is stronger than the redshifted emission in the cavity wall (discussed later) also support this interpretation. At the position of VLA 1623W, the blueshifted outflow lobe is distorted. We note that C$^{18}$O condensation is also seen on the northwestern side of VLA 1623W, the location of which is close to the position of Knot C identified in \cite{bontemps97}. 

Figure \ref{pv13coc18o} shows the PV diagram cut along the positions of VLA 1623B, VLA 1623W, and Knot C detected in \cite{bontemps97}. Significant blueshifted emission is seen along the cavity-wall. The highest velocity component is located at the position of VLA 1623B in $^{13}$CO and C$^{18}$O emission, and the velocity widths (FWHM) at the position of VLA 1623B are 5.9 km s$^{-1}$ and 4.2 km s$^{-1}$, respectively. The next high-velocity component in blueshifted emission is associated with VLA 1623W, and the velocity widths (FWHM) there are 4.3 km s$^{-1}$ and is 3.9 km s$^{-1}$, respectively. At the position of the Knot C, no high-velocity component exists, however the velocity width (FWHM) is 2.5 km s$^{-1}$ in C$^{18}$O. This width is 2.5 times larger than the average velocity dispersion in Oph-A core of 0.43 km s$^{-1}$ (corresponding to FWHM of 1.0 km s$^{-1}$) calculated from the results of single-dish observations of the C$^{18}$O(3--2) line \citep{white14}.  

\subsection{$^{12}$CO excitation temperatures}
In Figures \ref{peak} (a) and (b), we plot the peak temperature map of $^{12}$CO on the C$^{18}$O total integrated intensity map. 
At the position of VLA 1623B, $^{12}$CO emission exhibits the highest temperature in the map at 89.3 K $\pm$0.4 K. This value is higher than $\sim$60 K found by \cite{maury12} as the peak temperature of $^{12}$CO(2--1) obtained by SMA observations with a spectral resolution of 0.5 km s$^{-1}$ and the very extended configuration, giving the 0\farcs4 beam. The good $uv$ coverage of our observations likely resolve out less emission than those of \citep{maury12}. Assuming that $^{12}$CO is optically thick, the peak temperature corresponds to the physical temperature, or is a lower limit to it in the presence of beam dilution. The positions of VLA 1623W and Knot C show higher temperatures of 35--40 K compared to the area inside the outflow lobes of $\sim$30 K. At the positions of both VLA 1623B/W, the high-temperature areas are elongated along the PAs of the dust continuum sources identified by \cite{harris18}, as shown in Figures \ref{peak} (b). 

We estimate the excitation temperatures considering the optical depths calculated from $^{12}$CO , $^{13}$CO and C$^{18}$O spectra, as these will provide more precise values. The spectra at the positions of VLA 1623 A/B/W, and Knot C are shown in Figure \ref{spectrum}. The optical depths and excitation temperatures are calculated using the following equations, assuming the abundance ratios between $^{12}$CO and $^{13}$CO, and $^{13}$CO and C$^{18}$O are 71 and $\sim$6, respectively \citep{Frerking87}; 
\begin{eqnarray}
\frac{T^*_{\rm R}({\rm ^{12}CO})}{T^*_{\rm R}({\rm ^{13}CO})}=\frac{1-\exp[-\tau({\rm ^{12}CO})]}{1-\exp[-\tau({\rm ^{12CO})/71]}},\nonumber\\
\frac{T^*_{\rm R}({\rm ^{13}CO})}{T^*_{\rm R}({\rm C^{18}O})}=\frac{1-\exp[-\tau({\rm ^{13}CO})]}{1-\exp[-\tau({\rm ^{13}CO})/6]},
\label{eq:opt}
\end{eqnarray}
where $T^*_{\rm R}({\rm ^{12}CO})$, $T^*_{\rm R}({\rm ^{13}CO})$, and $T^*_{\rm R}({\rm C^{18}O})$ are the radiation temperatures and $\tau({\rm ^{12}CO})$ and $\tau({\rm ^{13}CO})$ are the optical depths of each line. We calculated optical depths within the velocity ranges shown in Figure \ref{spectrum}, i.e., where the emission of both lines is detected above $5\sigma$. We avoid the velocity ranges where spectra exhibit self-absorptions. The excitation temperatures of $^{12}$CO and $^{13}$CO, i.e., $T_{\rm ex}({\rm ^{12}CO})$ and $T_{\rm ex}({\rm ^{13}CO})$, are calculated from the following equations; 
\begin{eqnarray}
T_{\rm ex}({\rm ^{12}CO}) = T^*_{\rm R}({\rm ^{12}CO})/[1-\exp(-\tau^{\rm 12CO})]+T_{\rm bg}, \nonumber\\
T_{\rm ex}({\rm ^{13}CO}) = T^*_{\rm R}({\rm ^{13}CO})/[1-\exp(-\tau^{\rm 13CO})]+T_{\rm bg}
\label{eq:opt}
\end{eqnarray}
where $T_{\rm bg}=2.7$ K is the temperature of the cosmic microwave background radiation. The resulting values are summarized in Table \ref{tbl:spec}. The excitation temperatures of $^{12}$CO calculated at the positions of VLA 1623W and Knot C ($T_{\rm ex}\sim$35 K) are higher than $\sim26$ K, the averaged value integrated in SB2. All values shown in Table \ref{tbl:spec} are lower limits since all spectra exhibit self-absorption or the effects of resolved-out emission at the systemic velocity of $V_{\rm LSR}$=4 km s$^{-1}$. 

\section{DISCUSSION} \label{sec:dis}
\subsection{Are VLA 1623B and VLA 1623W knots or protostars?} \label{sec:dis:bw}
From our observations, we find twin outflows launched from the VLA 1623A/B region. The driving source of these outflows could be two of the three dust continuum sources, i.e., VLA 1623Aab and VLA 1623B. The latter source, however, is dynamically interacting with outflow-2 from the evidences listed below. Thus, we conclude that the both outflows are ejected from VLA 1623Aab as discussed later. 
\begin{itemize}
\item VLA 1623W is located on the blueshifted outflow cavity wall traced by C$^{18}$O emission, and a core-like structure is associated with VLA 1623W. The blueshifted outflow lobe is distorted at the position of VLA 1623W, indicating that VLA 1623W is affected by outflow-2.
\item The velocity widths of blueshifted $^{13}$CO and C$^{18}$O emission exhibit the largest values at the positions of VLA 1623B/W, i.e., 4.3--5.9 km s$^{-1}$ in $^{13}$CO and 3.9--4.2 km s$^{-1}$ in C$^{18}$O compared to the typical FWHM of C$^{18}$O emission in Oph-A core, 1.0 km s$^{-1}$ \citep{white14}.
\item The brightness temperature of $^{12}$CO emission is 89.3$\pm$0.4 K at the position of VLA 1623B, a value higher than that obtained in a previous study \citep{maury12}. At the positions of VLA 1623W and Knot C, the $^{12}$CO brightness temperature of 35--40 K is higher than that obtained inside the outflow lobes. 
\end{itemize}
The sudden increase of the temperature and the velocity width at the position of VLA 1623B/W likely reflects gas heating due to outflow shocks \citep{bachiller96,hirano01}. Hence, both VLA 1623B/W are dynamically interacting with outflow-2. Knot C likewise has a nature similar to that of VLA 1623W, i.e., a local increase of brightness temperature, association with a C$^{18}$O condensation, and location inside the outflow cavity-wall. This similarity suggests that VLA 1623W is a shocked cloudlet. VLA 1623B/W have the same nature, i.e., the increasing of velocity widths and the brightness temperature. Therefore, we conclude that VLA 1623B is also a cloudlet shocked by outflow-2. 

Although some previous studies have claimed that VLA 1623B/W are Class-0 and Class-1 protostellar candidates \citep{looney00,murilo13-2}, our results support the scenario that VLA 1623B/W are shocked cloudlets, as suggested by \cite{bontemps97} and \cite{maury12}. \cite{heish19} proposed that the increase of velocity width around VLA 1623B is due to the shock of the accretion flows. Careful inspection is, however, required since our results show that the blueshifted C$^{18}$O emission is strongly affected from the molecular outflow and the origin of the blueshifted C$^{18}$O is not necessarily accretion flows. \cite{murilo13-2} performed SED fitting and estimated that the dust temperature is 5 K. Our result, which shows $^{12}$CO brightness temperature of $\sim$90 K, invalidates the existence of a cold core associated with VLA 1623B.   

If VLA 1623B is indeed a shocked cloudlet and not protostellar in nature, both outflow-1 and outflow-2 must be ejected from VLA 1623A, i.e., VLA 1623Aa and VLA 1623Ab. It may yet be possible that VLA 1623B/W are protostars surrounded by envelopes heated from the outflow-2 shock. Even so, both outflow-1 and outflow-2 are still likely launched from VLA 1623Aab. Outflow-1 is likely ejected from VLA 1623A since the PA of outflow-1 is perpendicular to the CB disk and the root of outflow-1 is located at VLA 1623A (see Figure \ref{coch_center} and Section \ref{sec:res:co1}). VLA 1623A is considered to be also the driving source of outflow-2, since outflow-2 interacts with VLA 1623B and cannot be driven by it. It is beyond the scope of this study to probe whether or not a protostar exists inside the shock heated envelope around VLA 1623B. 

\subsection{The configuration of outflows and disk misalignment of VLA 1623A} \label{sec:dis:mis}
As discussed in the previous section, two molecular outflows are likely being ejected from VLA 1623Aa and VLA 1623Ab. The blue- and redshifted lobes of outflow-1 and outflow-2 are located on opposite sides of VLA 1623A, i.e., the northwestern redshifted lobe and southeastern blueshifted lobe of outflow-1 and the northwestern blueshifted lobe and southeastern redshifted lobe of outflow-2. These outflows are likely being ejected in opposite directions across the plane of sky. Assuming that the molecular outflows are launched perpendicular to the disks, the direction of the disk rotational axes should be in the opposite direction across the plane of the sky. \cite{harris18} suggested that the inclination angles of the CB disk and CS disks around VLA 1623Aa and VLA 1623Ab are well aligned on the basis of their inclinations, $\sim55\arcdeg$, which were measured from the major- to minor-axis ratios of the dust continuum sources at $\sim0\farcs1$ resolution. The disk inclinations derived from the aspect ratios, however, provide us with two possibilities, i.e., the disk's northwest side corresponds to either its near or the far side. This uncertainty can be resolved with the observed outflow orientations. Here, we introduce a disk inclination angle taking a value of 0 to 180 degrees. Such definition of the inclination provides us with two possible cases $i_{\rm disk}$ and (180$\arcdeg-i_{\rm disk}$) for the CS disks with the same major- to minor-axis ratios. The CB disk and the CS disk around outflow-1 driving source (hereafter, disk-1) is considered to be aligned, since the PA of outflow-1, 120\arcdeg, is almost perpendicular to the CB disk PA of 25$\arcdeg\pm4\arcdeg$ \citep{harris18}. Assuming that disk-1 is aligned with CB disk, the disk-1 inclination angle is estimated to be 55$\arcdeg$. The inclination of the secondary (the CS disk hosting driving source of outflow-2; hereafter, disk-2) is calculated to be $180\arcdeg - 55\arcdeg=125\arcdeg$, since all the CB disk and CS disks in VLA 1623A have similar major- to minor-axis ratio \citep{harris18}. Hence, the difference in the inclination between disk-1 and disk-2 is 70\arcdeg. The overall schematic view of the VLA1623Aab system is shown in  Figure \ref{schematic_view}. 

VLA 1623A has been posited to be an equal-mass binary system \citep{harris18,kawabe18} with an intermediate separation of 36 au (see below). We assume that the binary orbit is circular and parallel to the CB disk. Therefore, the true separation can be calculated geometrically from the apparent binary separation using the following equation, $a=a_{\rm app}[\cos^2{\phi}\sin^2{i}+\sin^2{\phi}]^{-1/2}$. Here, $a$ is the binary separation, $a_{\rm app}$ is the apparent binary separation, $\phi$ is the orbit phase angle, and $i$ is the inclination of the binary orbit. $\phi$ is calculated from the apparent angle between the two protostellar disks  by the equation, $\tan{\phi}=\tan{(\phi_{\rm app}\cos{i})}$. Using $a_{\rm app}=0\farcs23$ (corresponding to 32 au), $\phi_{\rm app}=142\arcdeg$ (corresponding to $\phi=155\arcdeg$) from the results of \cite{harris18}, and the inclination of 55$\arcdeg$ \citep{murilo13}, the true separation is calculated to be 36 au. If inclination has the error of $\pm25\arcdeg$, the true separation is expected to be 31--44 au. The uncertainty of the inclination is calculated from the half-opening angle of the outflow-1. If the inclination is larger than 80$\arcdeg$, outflow-1 should display blue- and red-shifted emission on both sides of VLA 1623A given outflow-1's half-opening angle of $10\arcdeg$ (Table \ref{tbl:co}). The separation is roughly consistent with the expected binary separation of 30 au which is calculated from the equation $r=(1.8$--$2.6)a$, which is predicted by \cite{artymowicz94} from the theoretical models. Here, $r$ is the radius of the inner edge of the CB disk and was measured to be 0\farcs4 by \cite{harris18}.


As described above, disk-1 is considered to be aligned with the CB disk, since the PA of outflow-1's axis is perpendicular to the PA of the CB disk. Outflow-1 is more prominent in terms of $^{12}$CO intensity compared to outflow-2 and exhibits a pair of significant jets. This fact supports the idea that disk-1 is aligned with the CB disk, since mass accretion expected to be more efficient onto whichever protostellar disk is better aligned with the CB disk. In this case, disk-2 is misaligned with disk-1 and CB disk, which is in parallel to the binary orbit.
We note that it is uncertain which disk, the disk-1 or -2 is associated to which source in the protostellar binary VLA1623 Aab.


Recently, \cite{brinch16} found that the disks are misaligned in the Class-I binary system with the intermediate separation, IRS 43 (binary separation is 75$\pm4$ au). VLA 1623A is similar to IRS 43 in this regard. Since VLA 1623A has been identified as Class-0 source and so had an age younger than 10$^4$ yrs \citep{evans09}, our results demonstrate that disks can be already misaligned in the earliest phases of proto-binary formation. 

There are several possible models to explain the misalignment of the VLA 1623A disks. First, the fragmentation of the parent core before the protostars formed may explain the misalignment \citep{offner10,bate13}, since such models predict disk misalignment. Second, recent theoretical studies have suggested that the misaligned disks result from disk fragmentation. For example, a third companion could cause the misalignment by tidal interaction \citep{matsumoto15}. Alternatively, a difference between the large-scale and small-scale angular momentum due to turbulence could also create misalignment, since the local angular momentum determines the direction of the circumstellar disks \citep{Tsukamoto13}. We note, however, that the classical models assuming single unstable disk fragmentation \citep{adams89,bonnell94} cannot explain the  VLA 1623A system since these models predict disk alignment. More complicated scenarios, including turbulence or third companion, are required. 

\subsection{Wiggle of jet from VLA 1623A} \label{sec:dis:wig}
In this section, we discuss the possible origin of the wiggle and compare the wiggle pattern to models. We only discuss JR1 since its wiggle is clearer compared to that of JB1. 

\subsubsection{Possible Origin of the Wiggle}
We assume that the jet is launched from the primary star of the binary for convenience of discussion in this section. There are three possible scenarios to explain the wiggle: 1) rotation of the outflow axis due to the orbital motion of the driving source, i.e., the primary star \citep{masciadri02}; 2) short-term precession due to the perturbation of the disk around the primary star produced by a periodic torque from the secondary star \citep{katz82,bate00}; and 3) long-term precession due to tidal interaction between the disk of the primary star and the secondary star \citep{terqum99,bate00}. A misalignment between the binary orbit axis and the circumstellar disk rotational axis is required in the second and third scenarios. Precession in the second scenario is also referred to as ``nodding motion'' in \cite{katz82} or ``wobbling'' in \cite{bate00} to discriminate it from the ``precession'' in the third scenario. Hereafter, we refer to these three scenarios as the ``orbital jet,'' ``wobbling,'' and ``precession'' models, respectively, for convenience. 

To test these models quantitatively, three physical quantities of the wiggle  should be considered, i.e., its time period, spatial amplitude, and amplitude in velocity.
In all three scenarios, the time period of the wiggle is related to the binary orbital period. We estimate the binary orbital period from the equation, $\tau_b^2=\frac{4\pi^2}{G(M_1+M_2)}a^3$. Here, $M_1+M_2$ and $a$ are the summed masses of the primary and secondary, and the binary separation, respectively. Assuming that the binary separation is 36 au (see Section \ref{sec:dis:mis}) and $M_1+M_2$ is 0.22$\pm0.02$ $M_\sun$ from \citet{murilo13}, the binary orbital period is calculated to be 460$\pm20$ yr. The binary separation has, however, the range of 31--44 au as discussed in Section \ref{sec:dis:mis}, taking the inclination uncertainty of $\pm25\arcdeg$ into account. Including this uncertainty, the binary orbital period has a range of 370--620 yrs. 

The wiggle time period ($\tau_w$) and the binary orbital period ($\tau_b$) are related by the following equations for the above three scenarios, respectively; 1) $\tau_w \sim \tau_b$ \citep{masciadri02}, 2) $\tau_w\sim\frac{1}{2}\tau_b$ \citep{katz82,bate00}, and 3) $\tau_w\sim\frac{1.06}{\sigma^{3/2}\cos{\theta}}\tau_b$ \citep{terqum99,masciadri02}. In the latter, $\sigma$ is the ratio between the radius ($r_0$) of the circular orbit and the CS disk radius ($R$), i.e., $R/r_0$, and $\theta$ is the precession half-opening angle. note that 3) assumes that the masses of the primary and secondary are the same \citep{masciadri02}. Using $R=11$ au, which is calculated from the size of the dust continuum source associated with VLA 1623Aa of 0\farcs16 \citep{harris18}, $r_0$=18 au, and $\theta$=0.4$\arcdeg$ (the precession angle is calculated in the next section, Section \ref{sec:dis:wig:pre}), the wiggle time period in case 3) is calculated to be $\sim 2 \tau_b$. The quantity $r_0$ is calculated from the equation, $r_0=\frac{M_1}{M_1+M_2}a=1/2a=18$ au assuming an equal-mass binary and a binary separation of 36 au (see Section \ref{sec:dis:mis}).

Therefore, the wiggle time period should be 460 yrs, 230 yrs, or 920 yrs for the ``orbital jet'', ``wobbling'', or ``precession'' scenarios, respectively, if we assume that the binary orbital period is 460 yrs. In the following next two sections, we compare the wiggle to the ``orbital jet'' \citep{masciadri02} and ``precession'' (or ``wobbling'') \citep{eisloffel96,wu09,hsieh16} scenarios to investigate which scenario best explains the wiggle in JR1. ''Wobbling'' and ``precession'' depict similar motions, but differ in timescales. We can, therefore, discriminate between  the ''wobbling'' and ``precession'' scenarios if we compare the wiggle pattern with the sinusoidal pattern expected from the precession scenario.

\subsubsection{Comparison with Precession Model}
\label{sec:dis:wig:pre}
We compare the wiggle of JR1 shown in Figure \ref{jr1} with the model to calculate its period and amplitude.  Assuming that the jet velocity is constant, and the wiggle pattern is sinusoidal, the model of the precession is described as follows \citep{eisloffel96,wu09,hsieh16};
\begin{eqnarray}
\label{eq:wig}
\left(
\begin{array}{c}
x \\
y
\end{array}
\right)
&=&
\left(
\begin{array}{c}
\alpha\cdot l\cdot\sin{(2\pi l/\lambda + \phi_0)} \\
l\cdot \cos{i_{\rm jet}}
\end{array}
\right),
\end{eqnarray}
\noindent
where $x$ and $y$ are the usual Cartesian coordinates perpendicular and parallel to the jet axis, $\alpha$ is the precession amplitude, $\lambda$ is the spatial period of the precession, $l$ is the distance from the source along the jet axis, $\phi_0$ is the initial phase at the source, and $i_{\rm jet}$ is the inclination angle of the jet axis from the plane of sky. 

First, we calculate the PA of JR1 and the position offset of the JR1 driving source at the time JR1 is launched from the current position of VLA 1623A with a $\chi^2$ linear regression as a function of the distance from VLA 1623A. We attempt the analysis toward the peak positions along JR1, which are calculated from Figure \ref{jr1} (a), as shown in Figure \ref{jr1} (b). In this section, all fittings were attempted toward the data in 45\arcsec--70\arcsec offsets from VLA 1623A, since the pattern at 35\arcsec--45$\arcsec$ seems to be created by the outflow shock given the sudden increase of velocity width (observed Figure \ref{jr1} (c)). We also excluded the data at $>75\arcsec$, since the velocity decreases, unlike that seen at $<75\arcsec$ offsets. With this regression, we find the PA of JR1 and the position offset perpendicular to the jet axis in Figure \ref{jr1} (a) are -4.4$\pm$0.3$\arcdeg$ and 5\farcs0$\pm0\farcs3$, respectively. We note that Figure \ref{jr1} is rotated -30$\arcdeg$ while centered on the current position of VLA 1623A, meaning that the PA of JR1 is -64.4$\arcdeg$ in total. The best-fit linear function is shown in Figure \ref{jr1} (b). The offset of 5$\arcsec$ might be due to proper motion, since VLA 1623A is moving toward the south \citep{Sadavoy18}. 


We then calculate $\alpha$, $\lambda$, and $\phi_0$, using $\chi^2$ minimization with Equation \ref{eq:wig} toward the peak positions of JR1 along the outflow axis shown in Figure \ref{jr1} (b). We assumed that the PA is -64.4$\arcdeg$ (in total) and the position offset of the driving source from VLA 1623A is 5\farcs5, as calculated from the $\chi^2$ regression with the linear function. The inclination angle of the jet axis from the plane of sky ($i_{\rm jet}$) is expected to be 35$\arcdeg$, assuming that the jet is ejected perpendicular to the CB disk axis, which itself is inclined at 55$\arcdeg$ \citep{murilo13}. As a result, $\alpha$, $\lambda$, and $\phi_0$ are calculated to be (6.8$\pm0.4)\times10^{-3}$ (corresponding to the half precession opening angle of 0.39\arcdeg$\pm$0.02\arcdeg), 1360$\pm$10 au, and -120\arcdeg$\pm$30\arcdeg, respectively. The best-fit precession model is shown in Figures \ref{jr1} (a) and (b) with the cyan lines. The time period of the wiggle is calculated to be 180 yrs, assuming the axis of JR1 is inclined 35\arcdeg. For this estimate, we used the JR1 average velocity of $V_{\rm LSR}$=25 km s$^{-1}$ over the velocity range of $V_{\rm LSR}=+$22.0--$+$29.0 km s$^{-1}$ where the wiggle of JR1 is most clearly defined. We note that if the inclination has an uncertainty of $\pm 25\arcdeg$, the period has a range from 54 yrs to 530 yrs. This timescale is smaller than 960 yrs, timescale of the ``precession'' scenario. Hence, the ``wobbling'' scenario, which has a timescale of only 230 yrs, seems a more reasonable explanation of the observed behaviour. 

Though the ``wobbling'' scenario seems to explain the period of the wiggle, the velocity amplitude cannot be consistently explained by it. We calculate the velocity wiggle along the line-of-sight with the following equation in the precession model \citep{wu09,hsieh16};
\begin{equation}
\label{eq:wigvel}
V_{\rm LOS}=V_{\rm sys}\pm V_{\rm jet}[\cos{\theta}\sin{i_{\rm jet}}+\sin{\theta}\cos{i_{\rm jet}}\cos{(2\pi l/\lambda+\phi_0)}],
\end{equation}
where $V_{\rm LOS}$ is the line-of-sight velocity, $V_{\rm sys}$ is the systemic velocity, $V_{\rm jet}$ is the jet velocity, and $\theta$ is the half-opening angle of the wiggle. Using $\theta$=0.4\arcdeg, $\lambda$=1360 au, and $\phi_0$=-120$\arcdeg$, as estimated earlier (see Section \ref{sec:res:wig}), the expected velocity amplitude is 0.2 km s$^{-1}$ using $V_{\rm jet}$=37 km s$^{-1}$ (from $V_{\rm LSR}=21$ km s$^{-1}$ and $i_{\rm jet}$=35$\arcdeg$ in Section \ref{sec:dis:wig:pre}). This expected amplitude is smaller by a factor of $\sim$4 than the velocity wiggle of 0.9 km s$^{-1}$ that we detect in JR1. We plot the wiggle model obtained from the Equation \ref{eq:wigvel} in Figure \ref{jr1} (c). We assume that the jet is linearly accelerated and can describe using $V_{\rm jet}\sin{i_{\rm jet}}=a_{\rm acc,jet}\cdot y + V_{\rm init}$ (where $a_{\rm acc,jet}$ is the acceleration rate of the jet and $V_{\rm init}$ is the initial velocity). From the $\chi^2$ linear regression of the velocity pattern of JR1, $a_{acc,jet}$ and $v_{\rm init}$ are expected to be 0.3 km s$^{-1}$ arcsec$^{-1}$ and 4.9 km s$^{-1}$, respectively, as shown in Figure \ref{jr1} (c). Though the spatial period of the wiggle should realistically increase given the outflow acceleration, we just consider here a simple model with a fixed spatial period. Consequently, the amplitude of the wiggle in the velocity from the model is much smaller than the wiggle observed in the PV diagram.

\subsubsection{Comparison with Orbital Jet Model}
\label{sec:dis:wig:orb}
In this section, we compare the observed outflow behaviour with the ``orbital jet'' model. The spatial wiggle pattern expected from orbital motion is described using following equation \citep{masciadri02};  
\begin{eqnarray}
\label{eq:orb}
\left(
\begin{array}{c}
x \\
y
\end{array}
\right)
&=&
\left(
\begin{array}{c}
\mu\kappa z \sin{\left[\kappa\left( \frac{l}{r_0}\right) -\phi_0 \right]} + r_0\cos{\left[ \kappa\left( \frac{l}{r_0}\right) -\phi_0 \right]} \\
l\cdot \cos{i_{\rm jet}}
\end{array}
\right),
\end{eqnarray}
where $r_0$ is the orbital radius of the primary, $\kappa$ is the ratio between the jet velocity ($V_{\rm jet}$) and orbital velocity ($V_0$=$2\pi r_0 / \tau_b$ where $\tau_b$ is the binary orbital period) assuming a circular orbit and constant jet velocity. The other variables are the same as in Equation \ref{eq:wig}. We multiply here the equation in \cite{masciadri02} by an attenuation factor ($\mu$), since the spatial amplitude of the wiggle may be suppressed due to interaction with the circumstellar material. Hence, the velocity perpendicular to the jet should be also attenuated due to the interaction. Interestingly, the data show that the spatial and velocity amplitudes between 50$\arcsec$ and 60$\arcsec$ are larger than the amplitudes between 60$\arcsec$ and 70$\arcsec$. Although we need to take projection into account, we just consider it along the jet axis to simplify the situation. Notably, the effect of projection is dominant when the inclination of the jet axis is large. With these assumptions, we find the best fit binary orbital period ($\tau_b$), attenuation factor ($\mu$), and initial phase ($\phi_0$) are 174$\pm$2 yrs, 0.078$\pm$0.004, and 150$\arcdeg\pm30\arcdeg$, respectively. For these estimates, we fixed the jet velocity $V_{\rm LSR}(=V_{\rm jet}\sin{i_{\rm jet}}-V_{\rm sys}$) at 25 km s$^{-1}$ (see Section \ref{sec:res:wig}), the jet inclination ($T_{\rm jet}$) at 35$\arcdeg$, and the circular orbital radius at $r_0=1/2a\sim18$ au ($a$ is the binary separation), under the assumption of an equal-mass binary system \citep{harris18,kawabe18}. We also assumed that the PA of JR1 is -64.6$\arcdeg$ and the position offset of the driving source from the current position of VLA 1623A along the declination axis is 5\farcs5, as obtained in Section \ref{sec:res:wig}. The best-fit model is plotted in Figures \ref{jr1} (a) and (b). If we take the inclination uncertainty of 25$\arcdeg$ into account, the period is calculated to be smaller than 400 yrs. Hence, the period is roughly consistent with the binary orbital period of 370--620 yrs calculated above. We also plot the orbital jet model assuming no attenuation (i.e., $\mu=1$) in Figures \ref{jr1} (a) and (b). The expected amplitude is about ten times larger than that observed. Therefore, attenuation is required to explain the wiggle with the orbital jet model. We note that the attenuation factor cannot be taken into account for precession models, since the wiggle in the velocity in the precession model discussed in Section \ref{sec:dis:wig:pre} is just assumed to be the change of line-of-sight velocity, not the change of the true progression velocity.

The line-of-sight velocity of the jet expected with the wiggle from the orbital jet model is described with the following equation, since only the velocity structure perpendicular to the spatial pattern is observed due to the projection.
\begin{equation}
\label{eq:vorb}
V_{\rm LOS}=V_{\rm sys}\pm V_{\rm jet}\sin{i_{\rm jet}}\pm V_0\cos{\left[\kappa\left( \frac{l}{r_0}\right) -\phi_0 \right]}\cos{i_{\rm jet}}
\end{equation}
In Figure \ref{jr1} (c), we plot the expected behaviour using the same parameters calculated from fitting of the spatial pattern. To compare with the pattern in the velocity, $V_{\rm jet}\sin{i_{\rm jet}}$ in Equation \ref{eq:vorb} is assumed to be (0.3$\cdot x$+4.9) km s$^{-1}$ as we assumed in Equation \ref{eq:wigvel}. The resulting expected amplitude of wiggle-like structure in the velocity is $V_0=2.2$ km s$^{-1}$, about twice the amplitude of 0.9 km s$^{-1}$ observed in the PV diagram (Figure \ref{jr1} (b)). Our model, however, does not accommodate the deceleration of velocity perpendicular to the jet in the course of the jet propagation. We might need to  consider such an effect in the orbital jet model, because the suppression of velocity in the jet may occur because of interaction with circumstellar material. 

From our analysis, the orbital jet model seems to be the most reasonable explanation for the dominant mechanism of the wiggle since its predictions are roughly consistent with the amplitude of its wiggle-like structure in velocity. The orbital jet model, however, is insufficient to explain the spatial amplitude of the wiggle, because the predicted amplitude from that model is much larger than that observed (Figure \ref{jr1} (a)). Therefore, we cannot necessarily conclude that the observed wiggle is due to the orbital motion of the driving source. Mirror-symmetric (orbital jet) or point-symmetric (precession) behaviour is more general evidence that can help us to discriminate between the ``orbital jet'' and ``wobbling'' models \citep{hsieh16}. In the case of VLA 1623A, however, the wiggle of JB1 is not very clearly defined and is only detected at locations distant from VLA 1623A. Moreover, it is difficult to distinguish between mirror- or point-symmetric behaviour when the wiggle is not detected closer to the driving source. Observations at still higher spatial resolution will help us to identify the wiggle of blueshifted and redshifted jets closer to VLA 1623A and constrain better the dominant mechanism of the wiggle in its outflow. 

\section{Conclusion} \label{sec:con}
We have carried out ALMA observations toward the Class-0 protostar, VLA 1623 in the Ophiuchus molecular cloud ($d=137$ pc), in lines of $^{12}$CO, $^{13}$CO, and C$^{18}$O. The main results of these data are summarized as follows;
\begin{enumerate}
\item Our $^{12}$CO($J$=2--1) data reveal that the previously identified VLA 1623 outflow consists of twin spatially overlapped outflows. The PA of each outflow axis is different by roughly 10$\arcdeg$. Jet-like structures in each outflow lobe are also observed. 
\item Our $^{13}$CO and C$^{18}$O images reveal the detailed relations between dense gas associated with the two cm/mm sources, VLA 1623B and VLA 1623W, and the molecular outflow. The data strongly support the previously proposed idea that both VLA 1623B and VLA 1623W are shocked cloudlets as follows. The $^{13}$CO and C$^{18}$O results reveal that compact high-velocity components are associated with VLA 1623B/W, with velocity widths of 4.3--5.9 km s$^{-1}$ in $^{13}$CO and 3.9--4.2 km s$^{-1}$ in C$^{18}$O. VLA 1623W is inside the cavity-wall traced by C$^{18}$O emission. The peak brightness temperatures of $^{12}$CO emission at the positions of VLA 1623B/W are higher than at locations inside the outflow lobes. Particularly, the peak brightness temperature at the position of VLA 1623B is above 89 K.
\item The driving sources of the twin molecular outflows are identified as VLA 1623Aab. The outflow axes of the two molecular outflows are estimated to be inclined by 70$\arcdeg$ with respect to each other across the plane of sky, implying that the rotational axes of the binary system are also similarly tilted with respect to each other. VLA 1623A is one of the youngest known binary systems with largely misaligned circumstellar disks at a separation of 34 au. The initial conditions of binary formation is likely more complicated than those expected from disk fragmentation in quiescent environments. Thus, other effects, perhaps turbulence, must play roles in forming the system.
\item  The redshifted northwestern jet exhibits over three cycles of wiggle with a spatial period of 1360$\pm$10 au, corresponding to a time period of 180 yrs. Given the uncertainty of its inclination, the period is realistically in the range of 180--400 yrs. A wiggle-like structure in the velocity is also detected with an amplitude of 0.9 km s$^{-1}$. The velocity amplitude of the wiggle is consistent with the Kepler velocity (2.2 km s$^{-1}$) estimated for the VLA 1623Aab binary. Moreover, the upper range of the time period of the wiggle is close to the orbital period (460$\pm$20 yr) of the binary.
\end{enumerate}

\acknowledgments
We thank the anonymous referee for helpful comments. This paper makes use of the following ALMA data: ADS/JAO. ALMA\#2013.1.00839.S. ALMA is a partnership of ESO (representing its member states), NSF (USA), and NINS (Japan), together with NRC (Canada), MOST and ASIAA (Taiwan), and KASI (Republic of Korea), in cooperation with the Republic of Chile. The Joint ALMA Observatory is operated by ESO, AUI/NRAO, and NAOJ. Data analysis was in part carried out on the open-use data analysis computer system at the Astronomy Data Center (ADC) of the National Astronomical Observatory of Japan. N.H. acknowledges a grant from the Ministry of Science and Technology (MoST) of Taiwan (MoST 108-2112-M-001-017-, MoST 109-2112-M-001-023-). S.T. is supported by JSPS KAKENHI grant No. 18K03703. M.T. is supported by JSPS KAKENHI grant No.18H05442. K.T. is supported by JSPS KAKENHI grant No. 16H05998, 17KK0091, and 18H05440. This work was supported by JSPS KAKENHI Grant Number 13J10869.
\clearpage

\bibliography{vla1623}

\clearpage

\begin{deluxetable}{lccccc}[h]
\tabletypesize{\scriptsize}
\tablewidth{0pt}
\tablecolumns{5}
\tablecaption{Parameters for the ALMA observations\label{tbl:obs}}
\tablehead{
\colhead{} & \colhead{} & \colhead{} & \colhead{} & \colhead{} \\
\colhead{Line/Wavelength}  & \multicolumn{2}{c}{$^{12}$CO($J$=2--1)} & \colhead{$^{13}$CO($J$=2--1)}  & \colhead{C$^{18}$O($J$=2--1)}}
\startdata
Frequency(GHz) & \multicolumn{2}{c}{230.538} & 220.399 & 219.560 \\
Observation date & \multicolumn{4}{c}{2015 May 1 (12-m Array), 2014 aug 14-17 (7-m Array)} \\
Array Configuration & \multicolumn{4}{c}{12-m Array (40 ant, Minimum Baseline=12.5 k$\lambda$, Maximum Baseline=348 k$\lambda$), } \\
 & \multicolumn{4}{c}{7-m Array (10 ant, Minimum Baseline=8.1 k$\lambda$, Maximum Baseline=48 k$\lambda$)} \\
Channel Separation & \multicolumn{2}{c}{0.096 km s$^{-1}$} & 0.096 km s$^{-1}$ & 0.096 km s$^{-1}$ \\
Pointing Center & \multicolumn{4}{c}{$(\alpha_{\rm J2000}, \delta_{\rm J2000}$)=($16^{\rm h}27^{\rm m}26\fs506, -24\arcdeg31\arcmin28\farcs63$)}\\
On Source Time &  \multicolumn{4}{c}{33 min (12-m Array), 50 min (7-m Array)} \\
Bandpass Calibrators &  \multicolumn{4}{c}{J1517-2422 (12-m Array), J1733-1304 (7-m Array)} \\
Complex Gain Calibrator &  \multicolumn{4}{c}{J1625-2527}\\
Absolute Flux Calibrators & \multicolumn{4}{c}{Titan (12-m Array), Mars (7-m Array)} \\
Beam Size &  \multicolumn{2}{c}{1\farcs39$\times$0\farcs91 (PA=89.2\arcdeg)} & 1\farcs41$\times$0\farcs93 (PA=-81.0\arcdeg) & 1\farcs41$\times$0\farcs93 (PA=-81.1\arcdeg)\\
Map velocity grid & 1.0 km s$^{-1}$ & 0.2 km s$^{-1}$ & 0.2 km s$^{-1}$ & 0.2 km s$^{-1}$ \\
Map rms & 15 mJy beam$^{-1}$ & 50 m Jy beam$^{-1}$ & 34 mJy beam$^{-1}$ & 28 mJy beam$^{-1}$ \\
\enddata
\end{deluxetable}

\clearpage
\begin{deluxetable}{lcccccccc}[h]
\rotate
\tablewidth{0pt}
\tablecolumns{9}
\tablecaption{Outflow Parameters\label{tbl:co}}
\tabletypesize{\scriptsize}
\tablehead{
\colhead{ } & \colhead{ } & \colhead{ } & \colhead{ } & \colhead{ } & \colhead{ } & \colhead{ } & \colhead{ } & \colhead{ } \\
\colhead{ } & \colhead{Position Angle} & \colhead{Opening Angle} & \colhead{Velocity $^{\rm a}$} &\colhead{Total Integrated Intensity}& \colhead{Mass$^{\rm b}$} & \colhead{Energy$^{\rm c}$} & \colhead{Momentum$^{\rm c}$} & \colhead{Dynamical Timescale$^{\rm c}$} \\
\colhead{ } & \colhead{[$\deg$]} & \colhead{[$\deg$]} & \colhead{[km s$^{-1}$]} &\colhead{[Jy $\cdot$ km s$^{-1}$]}& \colhead{[$M_{\sun}$]} & \colhead{[$M_{\sun}\cdot$ (km s$^{-1})^2$]} & \colhead{[$M_{\sun} \cdot$ (km s$^{-1}$)]} & \colhead{[yr]}
}
 
\startdata
SB1 & 119 & 19 & -20.5$\sim$-0.5 & - & - & - & - & - \\
SR1 & -61 & 17 & 6.5$\sim$31.5 & - & - & - & - & - \\
SB2 & -50 & 20 & -14.5$\sim$-0.5 & - & - & - & - & - \\
SR2 & 128 & 31 & 6.5$\sim$24.5 & - & - & - & - & - \\
JB1 & 120 & - & -34.5$\sim$-10.5 & - & - &  - &  - & $>$0.8$\times10^{3}$\\
JR1 & -60 & - & 6.5$\sim$31.5 & - & - & - & - & $>$1.5$\times10^{3}$\\
JB2 & -48 & - & -7.5$\sim$-0.5 & - & - & - & - & $>$1.0$\times10^{3}$\\
JR2 & 136 & - & 6.5$\sim$13.5 & - & - & - & - & $>$2.8$\times10^{3}$\\
JB1 + SB1 & - & - & -34.5$\sim$-0.5 & $>$1.1$\times 10^{3}$ & $>$1.6$\times 10^{-3}$ & $>$2.8$\times10^{-2}$ & $>$32.8$\times10^{-2}$ & - \\
JR1 + SR1 & - & - & 6.5 $\sim$31.5 & $>$2.2$\times 10^{3}$ & $>$3.1$\times 10^{-3}$ & $>$4.2$\times10^{-2}$ & $>$42.3$\times10^{-2}$ & - \\
JB2 + SB2 & - & - & -14.5$\sim$-0.5 & $>$0.6$\times 10^{3}$ & $>$0.8$\times 10^{-3}$ & $>$0.8$\times10^{-2}$ & $>$4.1$\times10^{-2}$ & - \\
JR2 + SR2 & - & - & 6.5$\sim$23.5 & $>$0.1$\times 10^{3}$ & $>$0.2$\times 10^{-3}$ & $>$0.2$\times10^{-2}$ & $>$1.1$\times10^{-2}$ & - \\
\enddata
\tablenotetext{a}{The velocity is where emission is detected above the 5$\sigma$ level. The velocity channels of -0.5-6.5 km s$^{-1}$ are ignored since in these channels, $^{12}$CO(2--1) emission traces cloud components.}
\tablenotetext{b}{Masses are estimated on the assumption of optically thin emission, $T_{\rm ex}$=30 K, and $X[{\rm CO}]=10^{-4}$.}
\tablenotetext{c}{Values are estimated on the assumption of $i$=55\arcdeg\hspace{1ex}and $V_{\rm sys}=4.0$ km s$^{-1}$.}
\end{deluxetable}

\clearpage
\begin{deluxetable}{lllllllll}[h]
\rotate
\tablewidth{0pt}
\tablecolumns{9}
\tablecaption{Temperatures and Optical Depths of $^{12}$CO, $^{13}$CO \label{tbl:spec}}
\tabletypesize{\scriptsize}
\tablehead{
\colhead{ } & \colhead{ } & \colhead{ } & \colhead{ } & \colhead{ } & \colhead{ } & \colhead{ } & \colhead{ } & \colhead{ } \\
\colhead{ } & \colhead{Velocity Range ($^{12}$CO/$^{13}$CO)$^{\rm a}$} & \colhead{Velocity Range ($^{13}$CO/C$^{18}$O)$^{\rm b}$} & \colhead{$T_{12\rm CO}/T_{^13\rm CO}$} &\colhead{$T_{13\rm CO}/T_{{\rm C18O}}$}& \colhead{$\tau_{^{12\rm CO}}$} & \colhead{$\tau_{13\rm CO}$} & \colhead{$T_{\rm ex,12CO}$} & \colhead{$T_{\rm ex,13CO}$} \\
\colhead{ } & \colhead{$V_{\rm LSR}$ [km s$^{-1}$]} & \colhead{$V_{\rm LSR}$ [km s$^{-1}$]} & \colhead{ } &\colhead{ }& \colhead{ } & \colhead{ } & \colhead{[K]} & \colhead{[K]}}
\startdata
VLA 1623A & -1.1 to +0.7, +5.7 to +6.7  & +0.3 to +1.9, +5.3 to +6.3 & 5.3 ($\pm$0.1) & 4.7 ($\pm$0.3) & 14.7       & 0.5       & 64 & 50 \\
VLA 1623B  & -2.9 to +0.5, +5.7 to + 7.1 & -0.5 to +0.5, +4.9 to +5.5 & 8.3 ($\pm$0.1) & 4.1 ($\pm$0.2) & 9.1      & 0.9      & 80  & 30 \\
VLA 1623W  & -2.1 to +1.1                & -1.1 to +1.7, +4.5 to +5.1 & 3.8 ($\pm$0.1) & 4.0 ($\pm$0.2) & 21.5      & 1.0       & 35 & 47 \\
Knor C    & 1.1 to +1.3                  & \nodata                    & 11.5 ($\pm$2.1) & \nodata       & 6.4   & \nodata       & 35 & \nodata \\
outflow$^{\rm b}$ & -1.9 to +1.1, +5.7 to +6.7  & -0.9 to 1.7, +4.7 to +8.7 & 10.9($\pm$0.0) & 5.6($\pm$0.0) & 6.8   & 0.1       & 27  & 97 
\enddata
\tablenotetext{a}{We calculate the optical depths in the velocity ranges where both lines are detected above 5$\sigma$. The ranges are also avoid where the self absorptions are detected.}
\tablenotetext{b}{The values are calculated from the averaged value integrated in the area where the blueshifted lobe on NE side of VLA 1623A (SB2) is detected above 5$\sigma$ in the map shown in Figures \ref{13coc18o} (a) and (c).}
\end{deluxetable}

\clearpage

\clearpage

\clearpage

\clearpage
\begin{figure}
\epsscale{1.1}
\centering
\includegraphics[scale=0.9]{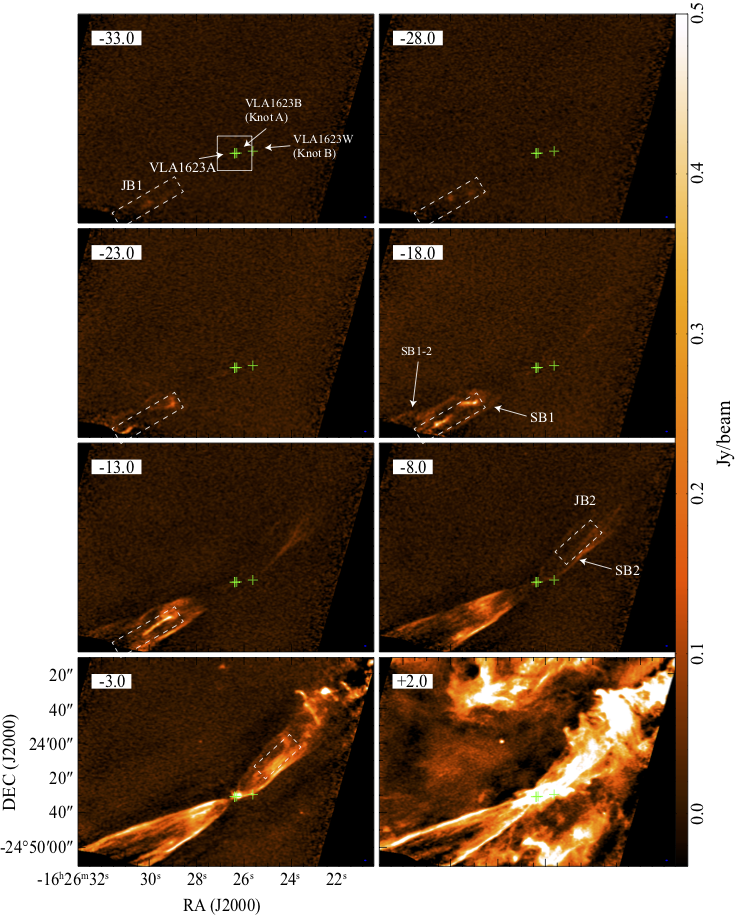}
\caption{Velocity channel maps of $^{12}$CO(2--1) emission within the velocity range from -33 to $+$32 km s$^{-1}$. The white box on the first velocity channel shows the area of Figure \ref{coch_center}. The symbols and dashed boxes are similar to those in Figure \ref{comom}. The solid box in the top-left panel depicts the area of Figure \ref{coch_center}.}
\label{coch}
\end{figure}

\addtocounter{figure}{-1}

\clearpage
\begin{figure}
\epsscale{1.1}
\centering
\includegraphics[scale=0.9]{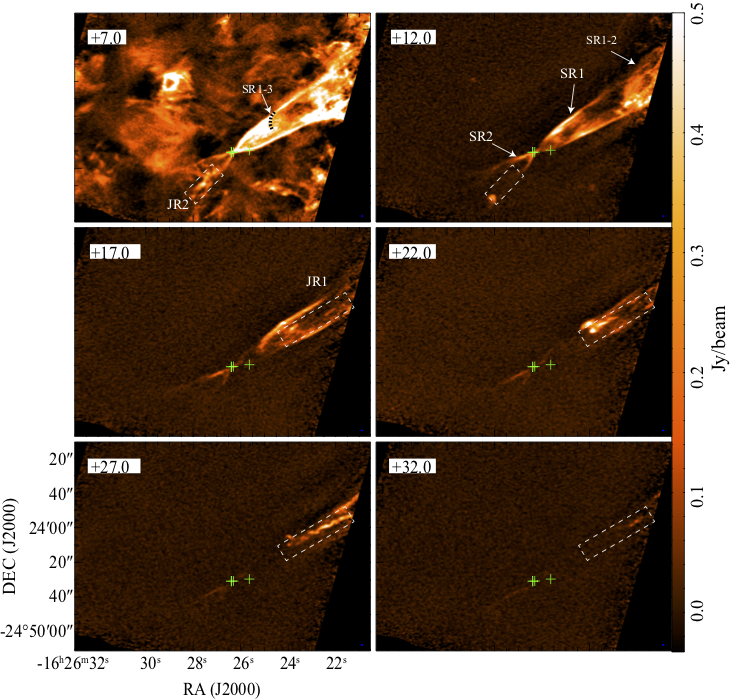}
\caption{Continued.}
\label{coch}
\end{figure}

\clearpage
\begin{figure}
\epsscale{1.0}
\centering
\includegraphics[scale=0.9]{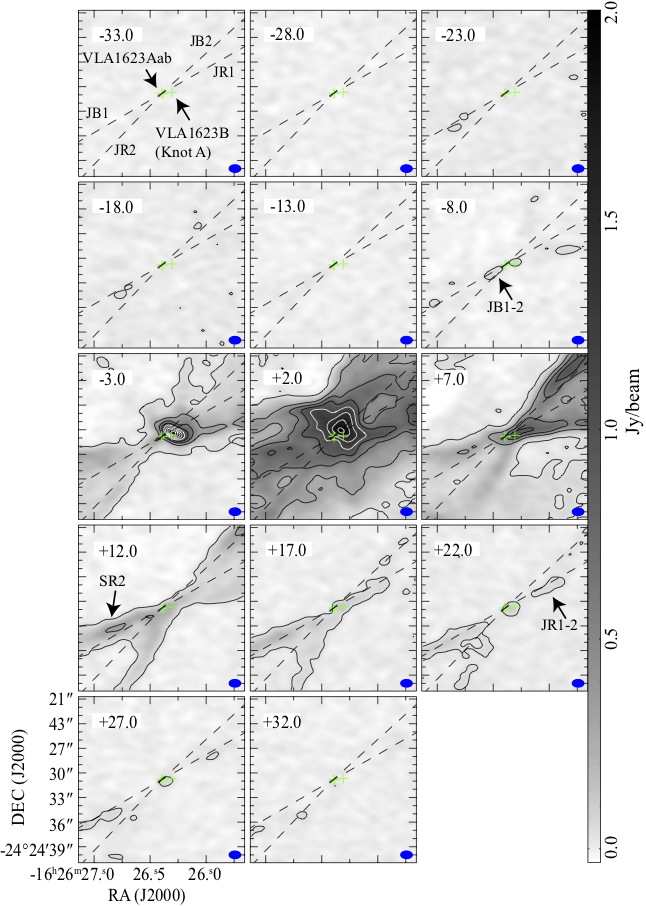}
\caption{Velocity channel maps of $^{12}$CO(2--1) for a 20$\arcsec\times20\arcsec$ area centered at VLA 1623A. Contour levels start at 5  $\sigma$ with an interval of 40 $\sigma$ (1 $\sigma$=0.0067 Jy beam$^{-1}$). The crosses show positions of VLA 1623Aab/B \citep{harris18}. Dashed lines are the same as that in Figure \ref{comom} (b).}
\label{coch_center}
\end{figure}

\clearpage
\begin{figure}
\epsscale{1.0}
\centering
\includegraphics[scale=0.3]{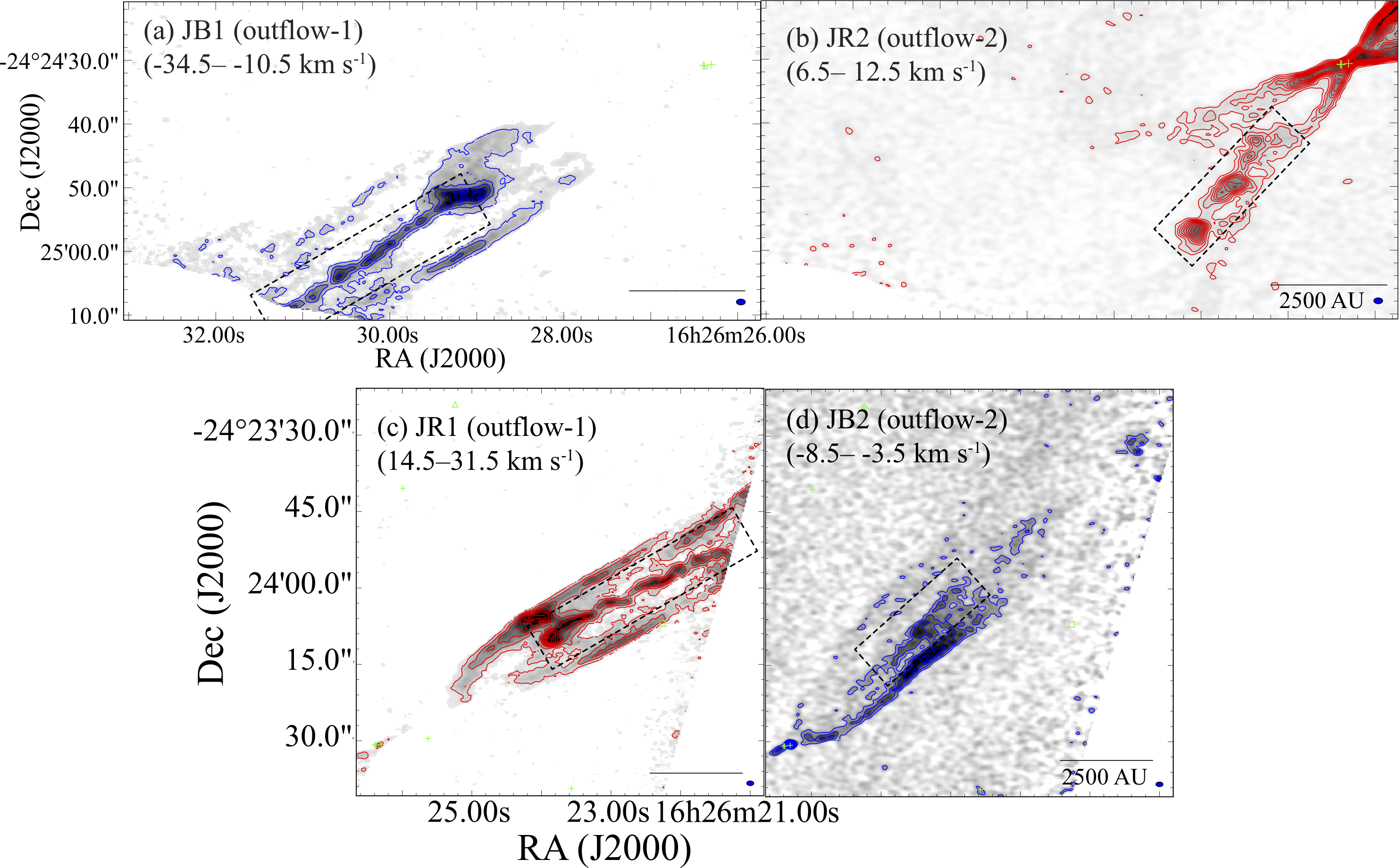}
\caption{The $^{12}$CO maps of the jets. The jets identified in Section \ref{sec:res:co1} and Figure \ref{coch} are depicted by dashed boxes. Each map is integrated over the velocity range described on top-left side of each panel. The contours start at 5  $\sigma$ with an interval of 10 $\sigma$ in (a) and (c), and at 5 $\sigma$ with an interval of 5 $\sigma$ in (b) and (d) ((a) 1 $\sigma$=0.07 Jy beam$^{-1}$ km s$^{-1}$, (b) 1 $\sigma$=0.03 Jy beam$^{-1}$ km s$^{-1}$, (c)  1 $\sigma$=0.06 Jy beam$^{-1}$ km s$^{-1}$, and (d) 1 $\sigma$=0.04 Jy beam$^{-1}$ km s$^{-1}$). The areas of the maps are the same as that shown in Figure \ref{subshells}.}
\label{jets}
\end{figure}

\clearpage
\begin{figure}
\epsscale{1.0}
\centering
\includegraphics[scale=0.3]{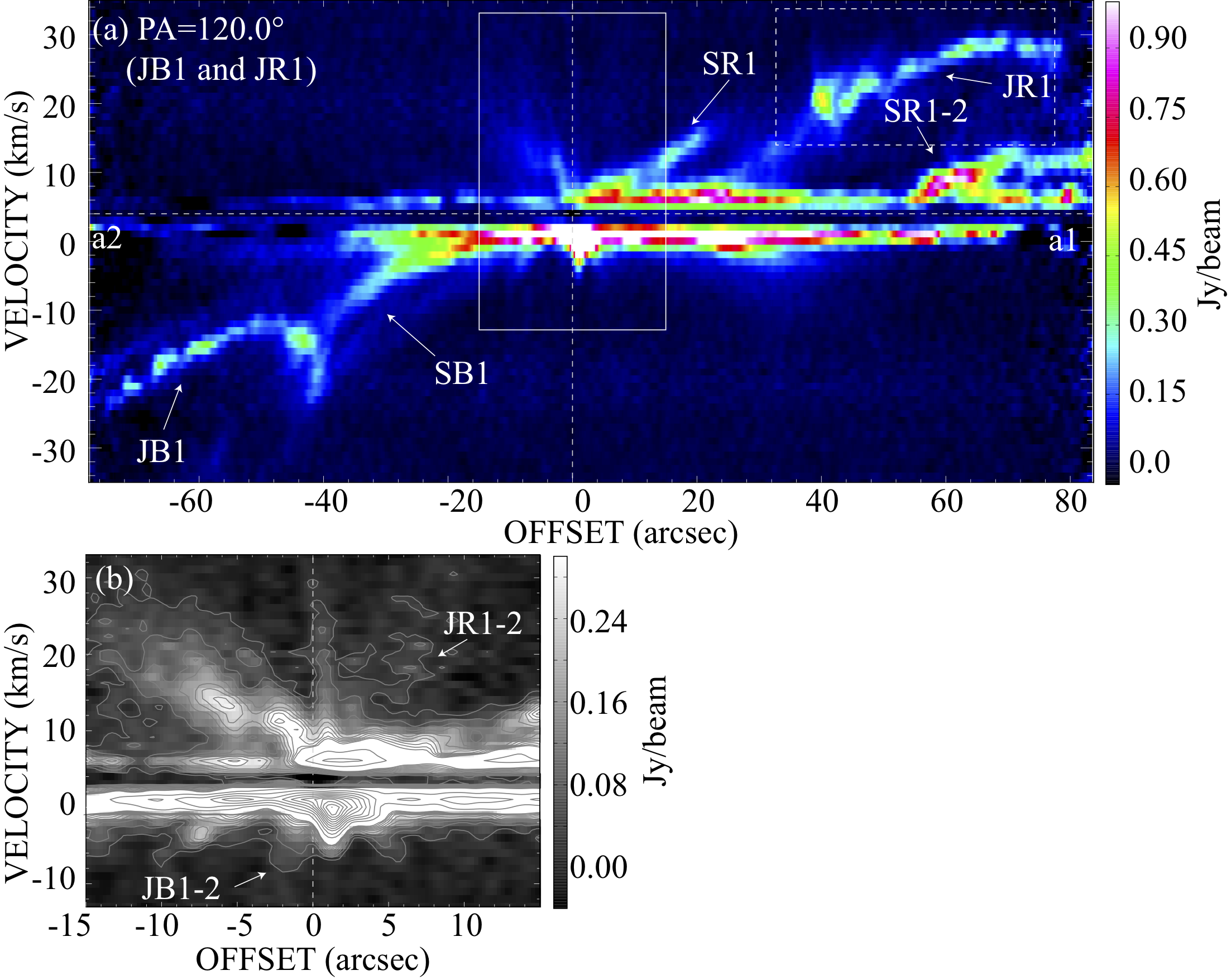}
\caption{(a) Position-velocity (PV) diagrams in $^{12}$CO(2--1) emission, cut along lines with a PA of 120.0\arcdeg (JB1 and JR1). (b) is the magnified view of (a). The contour levels in (b) start at 3 $\sigma$ and increase with an interval of 3 $\sigma$ until 30 $\sigma$ (1 $\sigma$=0.013 Jy beam$^{-1}$). After 30 $\sigma$, the contour intervals are 30 $\sigma$. In (a) and (b), the emission is averaged over 5$\arcsec$ and 1$\arcsec$ widths, respectively. The dashed lines in the vertical and horizontal directions show the position of VLA 1623Aa and the systemic velocity of $V_{\rm LSR}$=4 km s$^{-1}$. The solid and dashed boxes in (a) show the areas of (b) and Figure \ref{jr1} (c), respectively.}
\label{copv}
\end{figure}

\clearpage
\begin{figure}
\epsscale{0.8}
\centering
\includegraphics[scale=0.6]{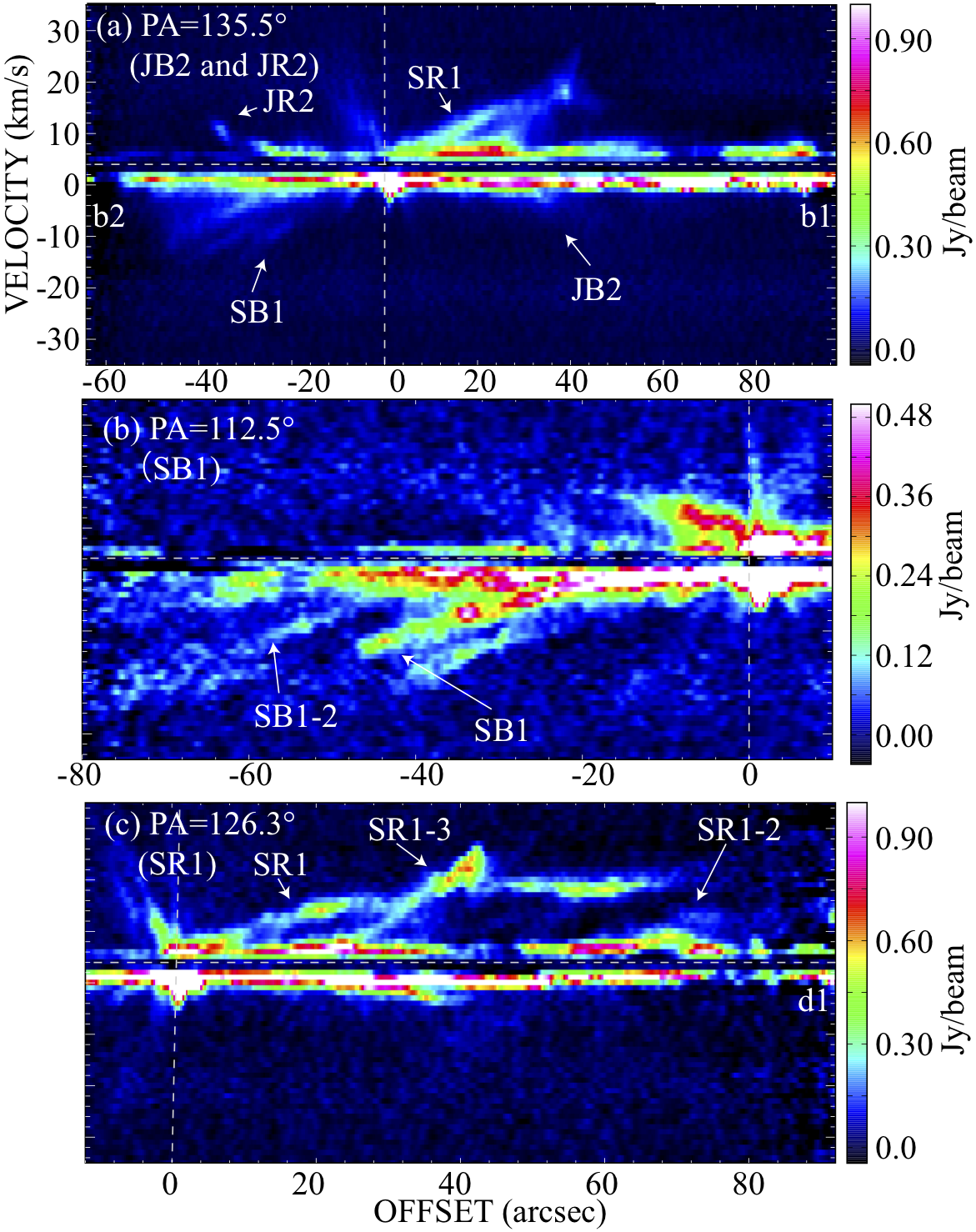}
\caption{PV diagrams in $^{12}$CO(2--1) emission, cut along lines with PAs of (a)135.5$\arcdeg$ (JR2 and JB2), (b)112.5$\arcdeg$ (SB1), and (c)126.3$\arcdeg$ (SR1). In (b), the emission is averaged over 8$\arcsec$ width. The dashed lines in the vertical and horizontal directions are the same as those in Figure \ref{copv}.}
\label{copv2}
\end{figure}

\clearpage
\begin{figure}
\epsscale{1.0}
\centering
\includegraphics[scale=0.3]{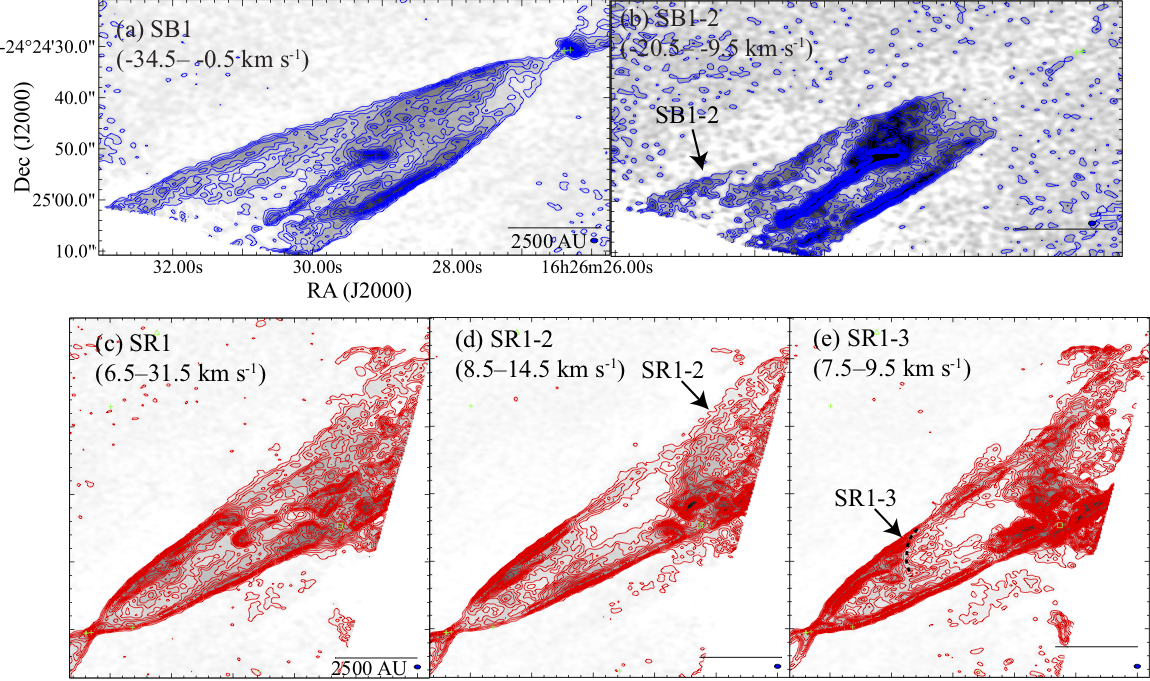}
\caption{Maps that show the internal structure of the lobes associated with outflow-1. The shells are identified in Section \ref{sec:res:copv} and Figure \ref{copv}. Each map is integrated over the velocity range described on top-left side of each panel. The contours in all Figures are start from 5 $\sigma$ and increase with an interval of 5 $\sigma$ ((a) 1 $\sigma$=0.09 Jy beam$^{-1}$ km s$^{-1}$, (b) 1 $\sigma$=0.05 Jy beam$^{-1}$ km s$^{-1}$, (c)  1 $\sigma$=0.08 Jy beam$^{-1}$ km s$^{-1}$, (d) 1 $\sigma$=0.04 Jy beam$^{-1}$ km s$^{-1}$, and (e) 1 $\sigma$=0.02 Jy beam$^{-1}$ km s$^{-1}$). The dashed curve in panel (e) is the SR1-3 identified in Section \ref{sec:res:copv}. Note that the areas of the panels (a) and (b) correspond to those of Figures \ref{jets} (a) and \ref{jets} (b), and the areas of panels (c), (d), and (e) correspond to those of Figures \ref{jets} (c) and \ref{jets} (d).}
\label{subshells}
\end{figure}

\clearpage
\begin{figure}
\epsscale{1.0}
\centering
\includegraphics[scale=0.6]{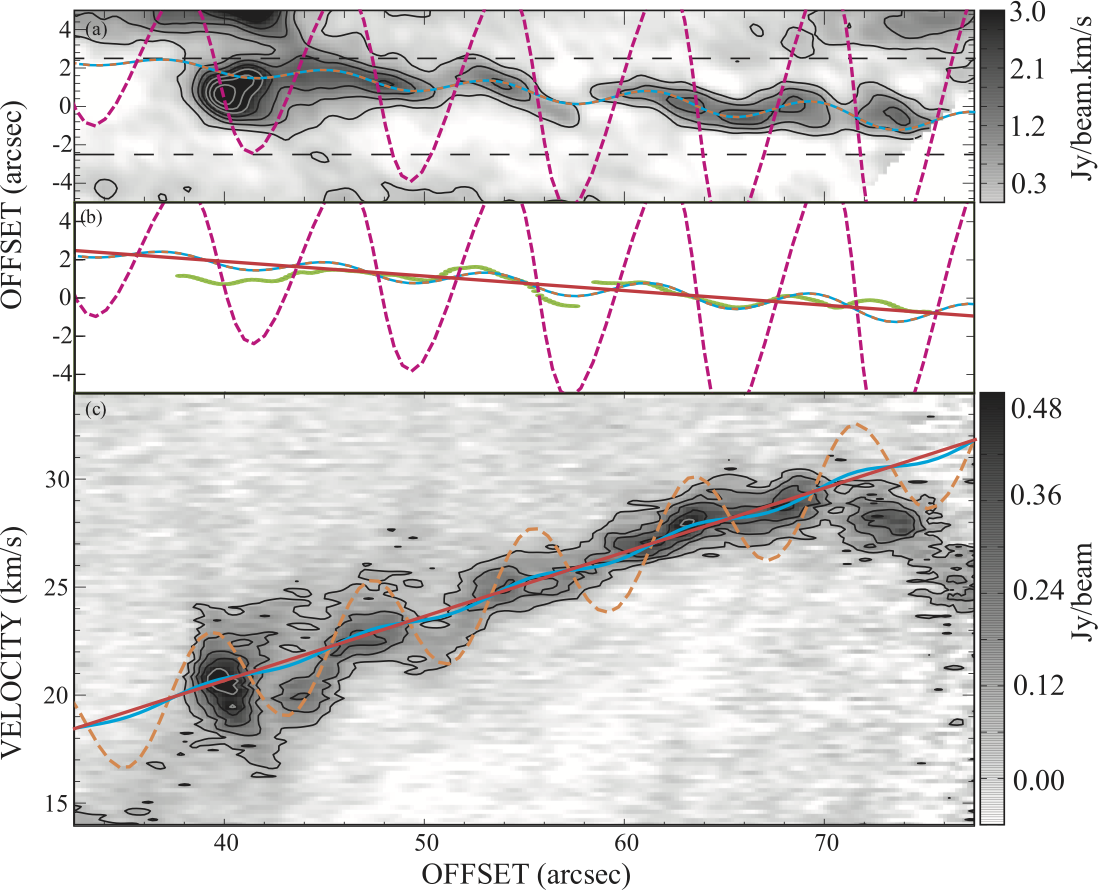}
\caption{Integrated intensity map of the redshifted jet, JR1, in $^{12}$CO(2--1) emission and the PV diagram cut along JR1 (PA=120$\arcdeg$). The map (a) is integrated over $V_{\rm LSR}=+$14.3 to $+$31.5 km s$^{-1}$ and rotated -30$\arcdeg$. The contour levels in (a) and (c) start at 5 $\sigma$ noise level with intervals of 5 $\sigma$ ((a) 1 $\sigma$=0.12 Jy beam$^{-1}$ km s$^{-1}$ and (c) 1 $\sigma$=0.02 Jy beam$^{-1}$). The images are created from the maps with a 0.2 km s$^{-1}$ velocity grid. Dashed lines in (a) show the area where panel (c) is created. The yellow-green curve in panel (b) shows the measured peak positions obtained from Gaussian fitting as a function of position offset along the JR1 axis calculated from (a). The cyan lines in (a), (b), and (c) show the best-fit precession model and orange dashed lines show the orbital jet model. Magenta dashed lines in (a) and (b) show the orbital jet model assuming no attenuation. The red line in (b) shows the best-fit linear function of the yellow-green plots. The red line in (c) shows the best-fit linear function of the P-V diagram.}
\label{jr1}
\end{figure}

\clearpage
\begin{figure}
\epsscale{1.0}
\centering
\includegraphics[scale=0.3]{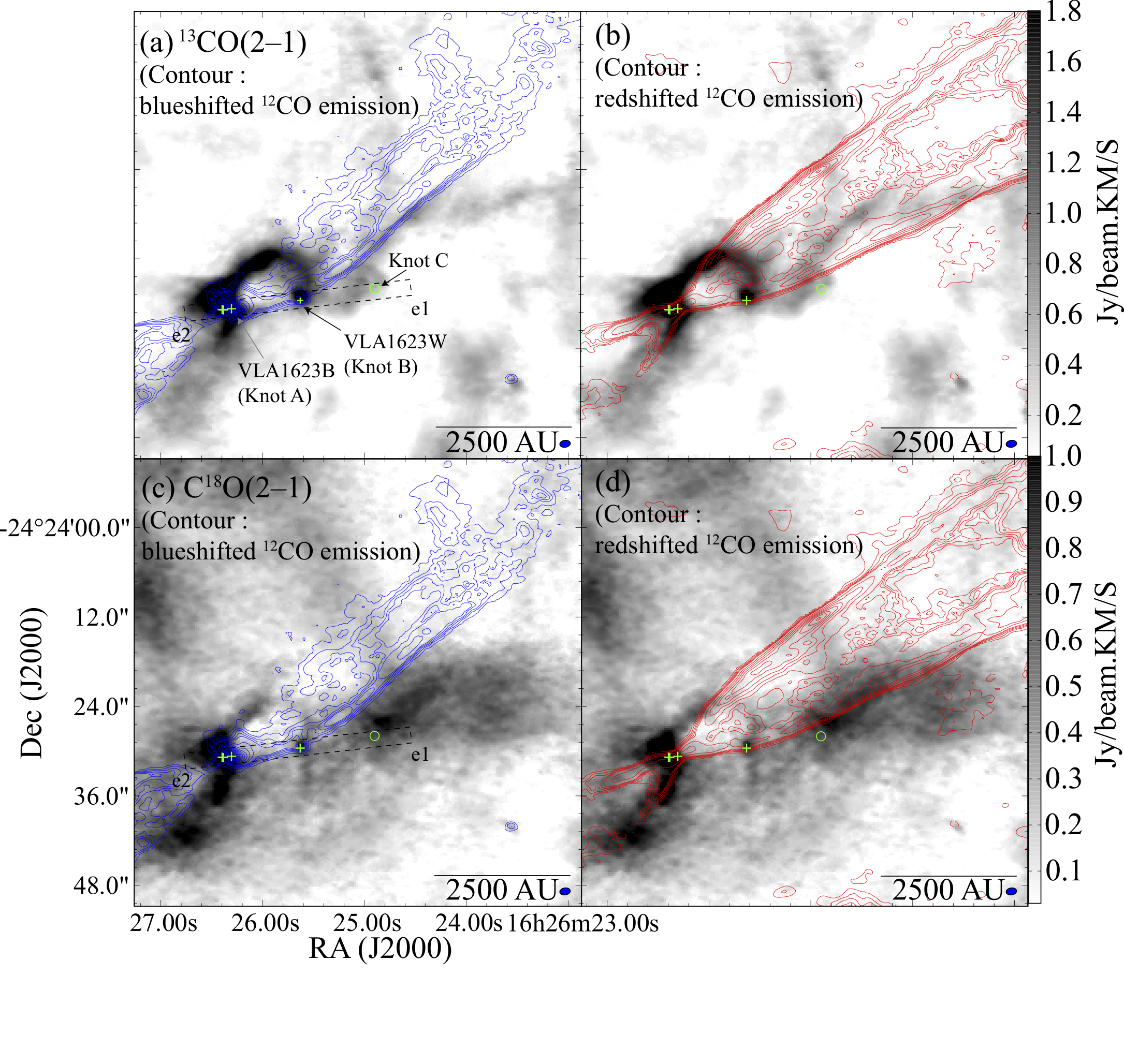}
\caption{$^{13}$CO and C$^{18}$O integrated intensity maps. Each map are integrated between (a) (b) $V_{\rm LSR}$=-2.3 to 7.7 km s$^{-1}$ and (c) (d) $V_{\rm LSR}$=-0.9 to 6.5 km s$^{-1}$. Blue contours on (a) (c) and red contours on (b) (d) show $^{12}$CO(2--1) blueshifted and redshifted intensities integrated between $V_{\rm LSR}$=-34.5 to -0.5 km s$^{-1}$ and $V_{\rm LSR}=+$6.5 to $+$31.5 km s$^{-1}$, respectively. Contour levels of blueshifted and redshifted integrated intensities start at 3 $\sigma$ and increase with an interval of 3 $\sigma$ (1 $\sigma$=0.09 Jy beam$^{-1}$ km s$^{-1}$ for blueshifted emission and 0.07 Jy beam$^{-1}\cdot$km s$^{-1}$ for redshifted emission, respectively) until 15 $\sigma$. After 15  $\sigma$, the contour interval is $10\sigma$. The symbols are the same as those shown in Figure \ref{coch}. The circle shows the position of Knot C identified in \cite{bontemps97}. Dashed boxes in the panels (a) and (c) show where the PV diagram in Figure \ref{pv13coc18o} is constructed.}
\label{13coc18o}
\end{figure}

\clearpage
\begin{figure}
\epsscale{1.0}
\centering
\includegraphics[scale=0.6]{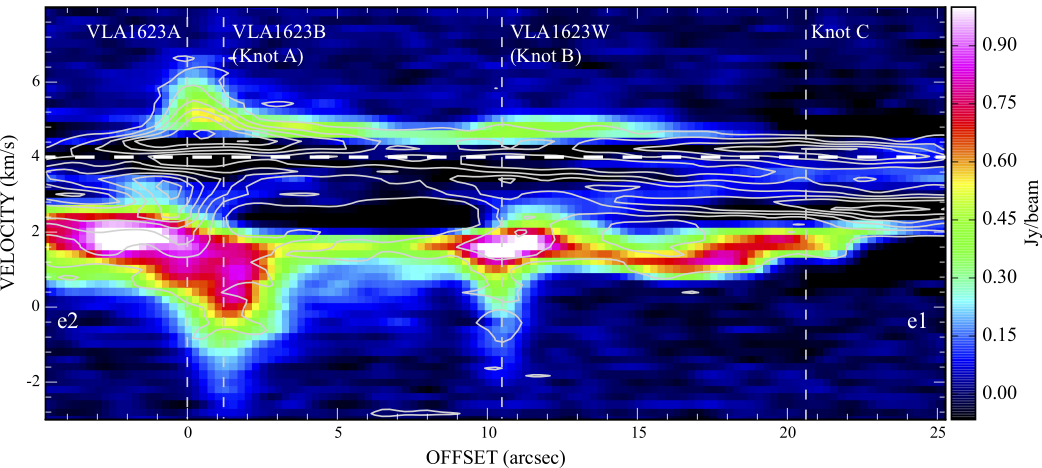}
\caption{Position-Velocity (PV) diagrams in $^{13}$CO(2--1) and C$^{18}$O(2--1), cut along lines with a PA of -83\arcdeg \hspace{1ex}. Background color shows $^{13}$CO(2--1) emission and the contour shows C$^{18}$O(2--1) emission. Contour levels start at 3 $\sigma$ and increase with an interval of 6 $\sigma$ (1 $\sigma$=0.015 Jy beam$^{-1}$). The dashed lines in the vertical and horizontal directions show the the positions of VLA 1623A, B (Knot A), W (Knot B), Knot C, and the systemic velocity, respectively.}
\label{pv13coc18o}
\end{figure}

\clearpage
\begin{figure}
\epsscale{1.0}
\centering
\includegraphics[scale=0.3]{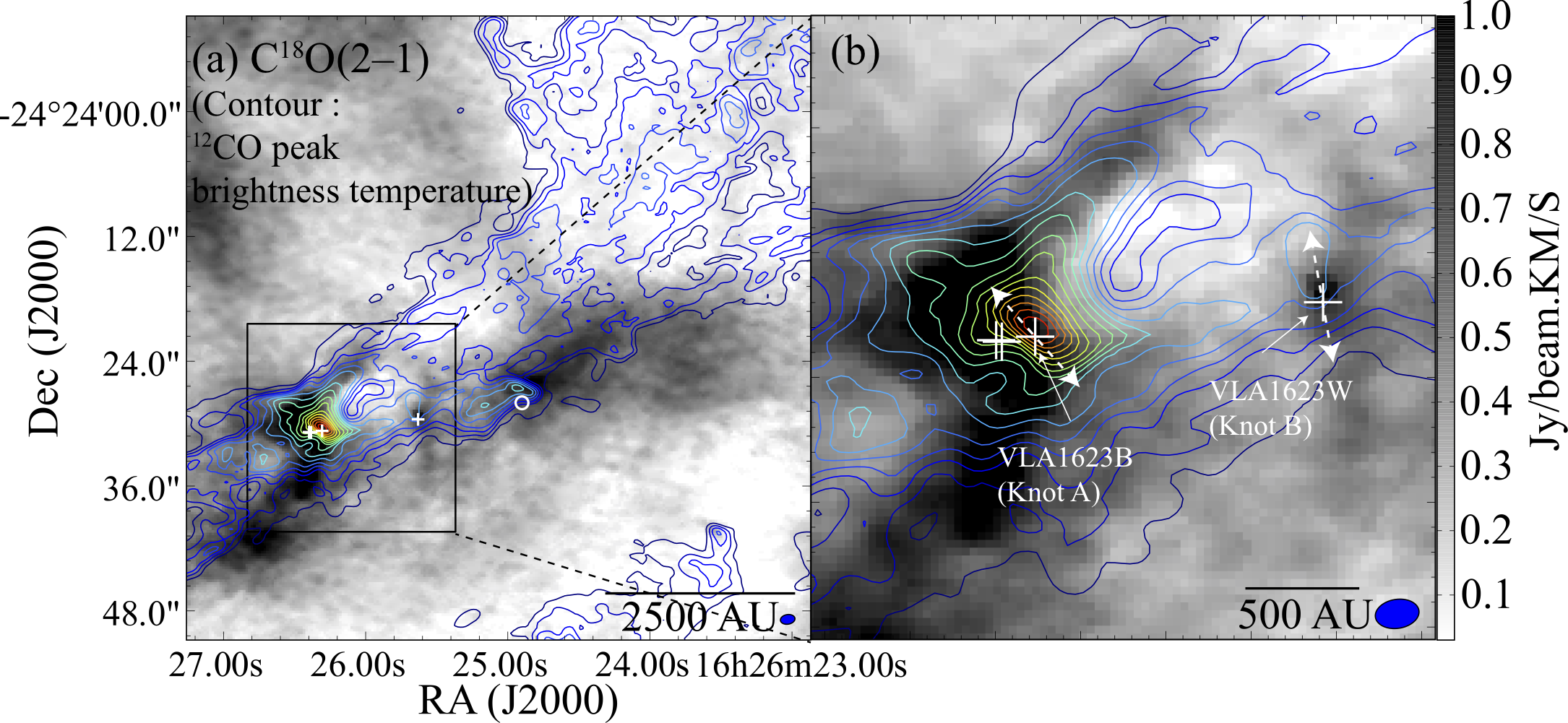}
\caption{$^{12}$CO peak brightness temperature map shown in contours and C$^{18}$O integrated intensity map shown in gray scale (left),  and their close-up (right). The gray scale and symbols are same as those shown in Figures \ref{13coc18o} (c) and (d). Contour levels start at 10 K and increase with an interval of 5 K. The two dashed white arrows in  panel (b) depict the two position angles of the elongations seen in the two dust continuum sources VLA1623B/W taken from \citep{harris18}.  }
\label{peak}
\end{figure}

\clearpage
\begin{figure}
\epsscale{1.0}
\centering
\includegraphics[scale=0.9]{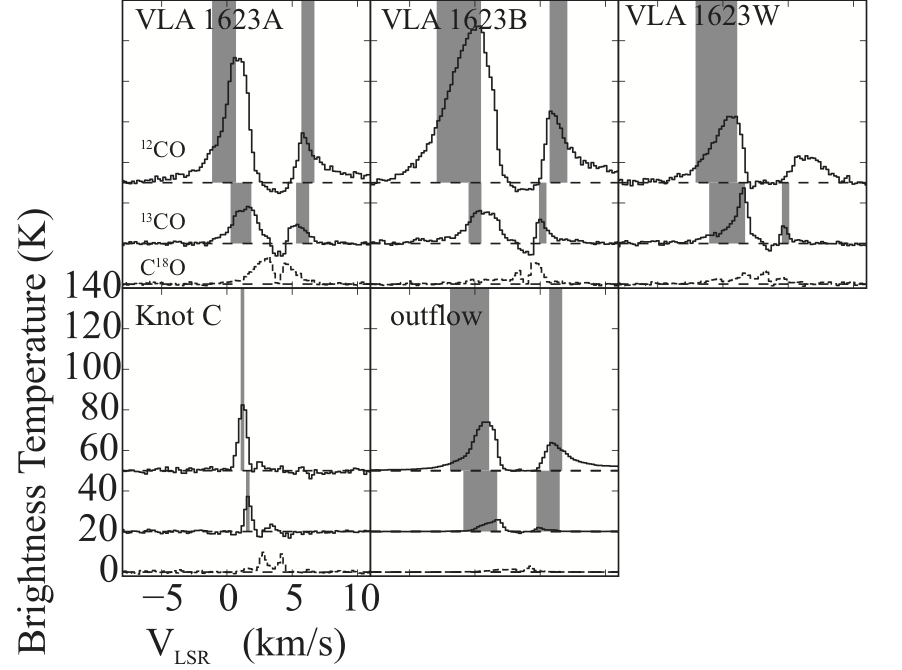}
\caption{The spectra of $^{12}$CO (top), $^{13}$CO (middle), and C$^{18}$O (bottom) at the positions of VLA 1623A/B/W, Knot C, and the outflow. The outflow spectra are calculated from the averaged value integrated over the area where the blueshifted lobe on the NE side of VLA 1623A (SB2) is detected above 5$\sigma$ in the map shown in Figures \ref{13coc18o} (a) and (c). The grey-filled regions indicate the velocity range where we calculate the optical depths shown in Table \ref{tbl:spec}. The $^{12}$CO spectra are created from the map with a 0.2 km s$^{-1}$ velocity grid.}
\label{spectrum}
\end{figure}

\clearpage
\begin{figure}
\epsscale{1.0}
\centering
\includegraphics[scale=0.5]{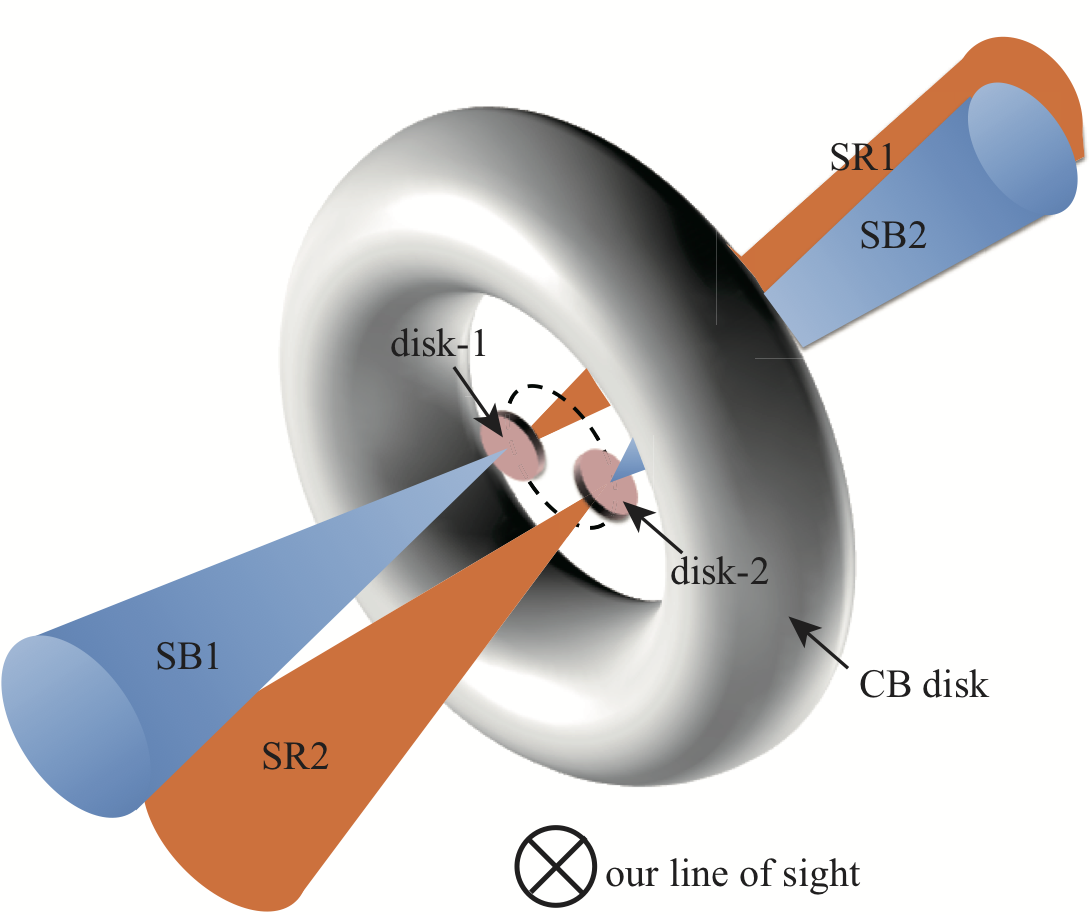}
\caption{The front view including each disk of the protostellar binary and its binary orbit, together with the CB disk.}
\label{schematic_view}
\end{figure}

\clearpage

\appendix
\section{$^{12}$CO(2--1) channel maps}
\label{sec:app:chan}
The $^{12}$CO(2--1) velocity channel maps with a velocity spacing of 2.0 km s$^{-1}$ are shown in Figure \ref{appcoch1}. The most powerful blueshifted jet, JB1, appears in the velocity range of $V_{\rm LSR}=-33.5$ to -9.5 km s$^{-1}$. The blueshifted outflow shell (SB1) associated with JB1 appears in the velocity range of $V_{\rm LSR}=$-19.5 to -1.5 km s$^{-1}$. A secondary blueshifted outflow shell inside SB1, we name SB1-2, apparent in the velocity range of $V_{\rm LSR}=$-18.5 km s$^{-1}$ to -10.5 km s${-1}$. SB1-2 is located across the southeastern side of SB1.

On the redshifted side, the outflow shells of SR1 and SR2 appear in the velocity channel of $V_{\rm LSR}=+$7.5 km s$^{-1}$. SR1 is apparent in the velocity range of $V_{\rm LSR}=$7.5 to 22.5 km s$^{-1}$, and SR2 is apparent in the velocity range of $V_{\rm LSR}=+$7.5 to $+$14.5 km s$^{-1}$. The jets, JR1 and JR2, are apparent in the velocity ranges of $V_{\rm LSR}=+$7.5 to $+$12.5 km s$^{-1}$ and $V_{\rm LSR}=+$16.5 to $+$30.5 km s$^{-1}$, respectively. The secondary outflow shell across SR1, which we name SR1-2 apparent in the velocity range of $V_{\rm LSR}=+$9.5 to $+$13.5 km s$^{-1}$. SR1-2 exists across SR1 on its northwestern side. The third outflow shell inside SR1, SR1-3, is apparent in the velocity range of $V_{\rm LSR}=+$13.5 to $+$17.5 km s$^{-1}$, and located in the northern part of SR1.

\renewcommand\thefigure{\thesection.\arabic{figure}}    
\setcounter{figure}{0}    

\begin{figure}
\epsscale{1.0}
\centering
\includegraphics[scale=0.9]{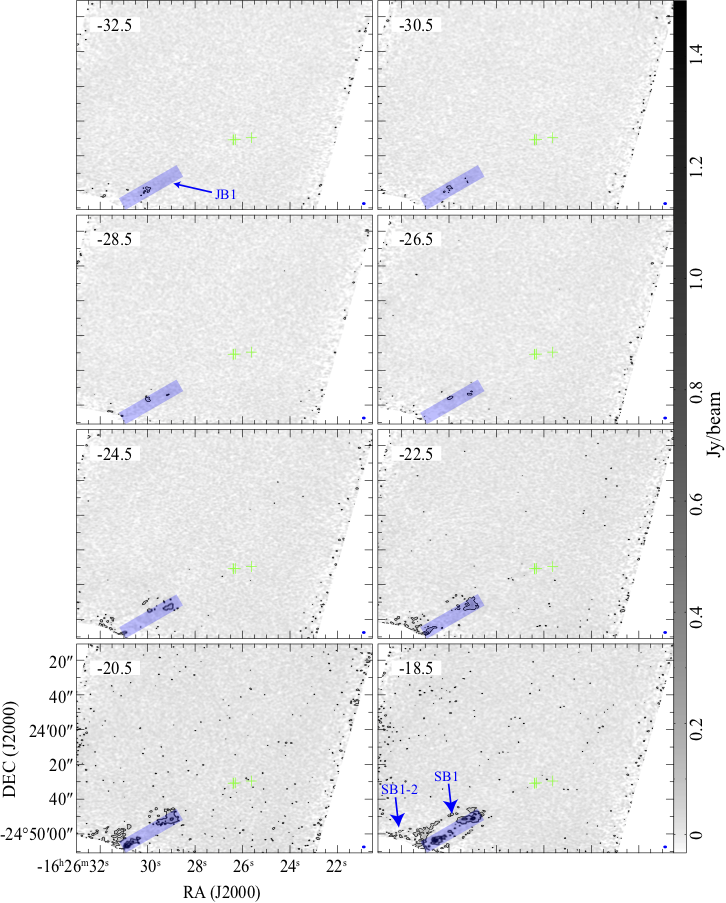}
\caption{Velocity channel maps of $^{12}$CO(2--1) for the velocity range of -33 to $+$30 km s$^{-1}$. The velocity spacing is 2.0 km s$^{-1}$. The symbols are the same as those defined in Figure \ref{coch}. The positions of jets (JB1, JB2, JR1, and JR2) are marked by blue and red translucent boxes. The contours start from 5 $\sigma$ and increase with an interval of 20 $\sigma$ where 1 $\sigma$=0.011 Jy beam$^{-1}$.}
\label{appcoch1}
\end{figure}

\clearpage
\addtocounter{figure}{-1}
\begin{figure}
\epsscale{1.0}
\centering
\includegraphics[scale=0.9]{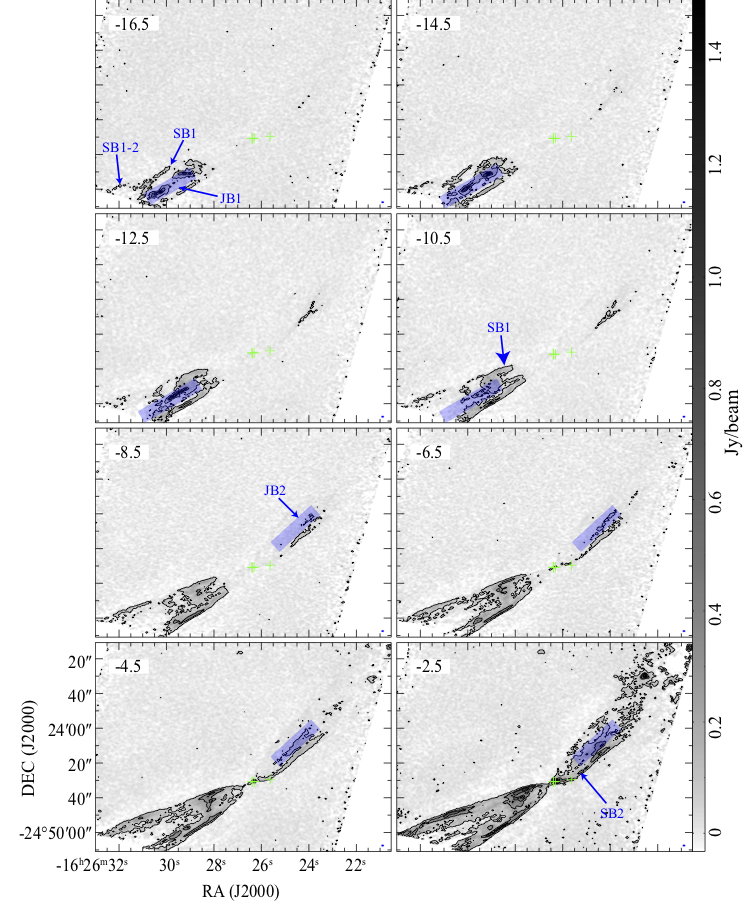}
\caption{Continued.}
\label{appcoch1}
\end{figure}

\clearpage
\addtocounter{figure}{-1}
\begin{figure}
\epsscale{1.0}
\centering
\includegraphics[scale=0.9]{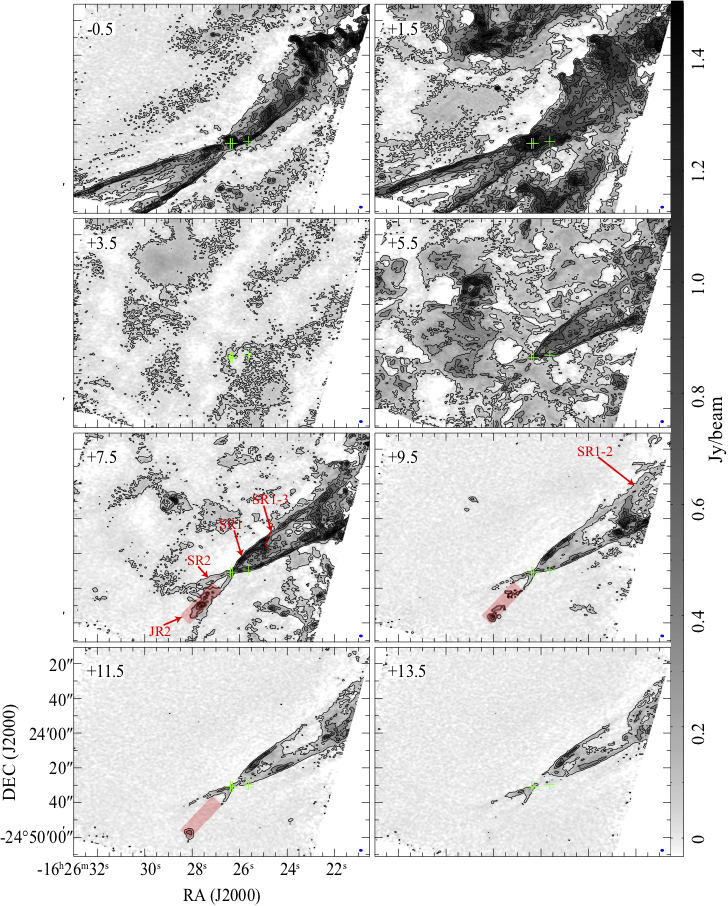}
\caption{Continued.}
\label{appcoch1}
\end{figure}

\clearpage
\addtocounter{figure}{-1}
\begin{figure}
\epsscale{1.0}
\centering
\includegraphics[scale=0.9]{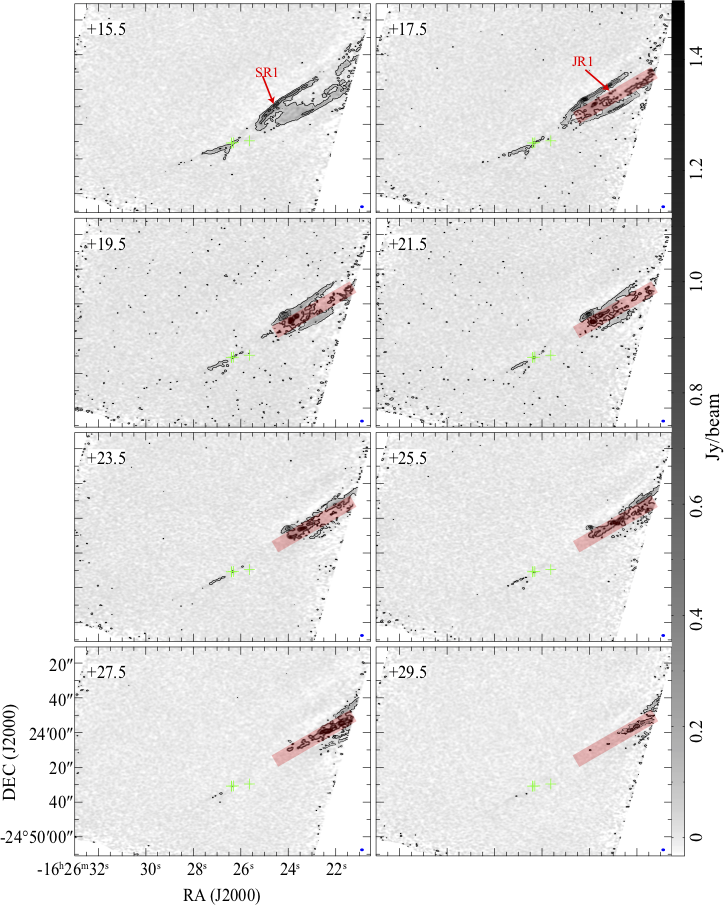}
\caption{Continued.}
\label{appcoch1}
\end{figure}



\end{document}